\documentclass[a4paper,fleqn,usenatbib,useAMS]{mnras}

\usepackage{graphicx}	
\usepackage{amsmath}	
\usepackage{amssymb}	
\usepackage{multicol}        
\usepackage{bm}		
\usepackage{pdflscape}	

\newcommand{\kms}{\,km\,s$^{-1}$} 

\usepackage{gensymb} 
\usepackage{soul} 

\newcommand{\teff}{T$_{\rm eff}$}
\newcommand{\logg}{$\log g$}

\def\kms{\,{\rm km~s^{-1}}}

\def\vlos{V_{\rm los}}

\def\ltsima{$\; \buildrel < \over \sim \;$}
\def\simlt{\lower.5ex\hbox{\ltsima}}
\def\gtsima{$\; \buildrel > \over \sim \;$}
\def\simgt{\lower.5ex\hbox{\gtsima}}

\def\ltsima{$\; \buildrel < \over \sim \;$}
\def\simlt{\lower.5ex\hbox{\ltsima}}
\def\gtsima{$\; \buildrel > \over \sim \;$}
\def\simgt{\lower.5ex\hbox{\gtsima}}

\title[The selection function of the RAVE survey]{The selection function of the RAVE survey}

\author[J.~Wojno et al.]{
Jennifer Wojno,$^{1}$\thanks{E-mail: jwojno@aip.de (JW)} 
Georges Kordopatis,$^{1}$
Tilmann Piffl,$^{2}$
James Binney,$^{2}$ 
\newauthor
Matthias Steinmetz,$^{1}$
Gal Matijevi\v c,$^{1}$
Joss Bland-Hawthorn,$^{3}$
Sanjib Sharma,$^{3}$
\newauthor
Paul McMillan,$^{4}$
Fred Watson,$^{5}$
Warren Reid,$^{6,7}$
Andrea Kunder,$^{1}$
Harry Enke,$^{1}$
\newauthor
Eva K.\ Grebel,$^{8}$
George Seabroke,$^{9}$
Rosemary F.\ G.\ Wyse,$^{10}$
Toma\v z Zwitter,$^{11}$ 
\newauthor
Olivier Bienaym\'{e},$^{12}$
Kenneth C. Freeman,$^{13}$
Brad K. Gibson,$^{14}$
Gerry Gilmore,$^{15}$
\newauthor
Amina Helmi,$^{16}$
Ulisse Munari,$^{17}$
Julio F.\ Navarro,$^{18}$
Quentin A. Parker,$^{19}$
\newauthor
Arnaud Siebert$^{12}$
\\
$^{1}$Leibniz Institut f\"ur Astrophysik Potsdam, An der Sternwarte 16, 14482 Potsdam, Germany\\
$^{2}$Rudolf Peierls Centre for Theoretical Physics, 1 Keble Road, Oxford OX1 3NP, UK\\
$^{3}$Sydney Institute for Astronomy, School of Physics A28, University of Sydney, NSW 2006, Australia \\
$^{4}$Lund Observatory, Lund University, Department of Astronomy and Theoretical Physics, Box
43, SE-22100 Lund, Sweden \\
$^{5}$Australian Astronomical Observatory, North Ryde, NSW 2113, Australia \\
$^{6}$Department of Physics and Astronomy, Macquarie University, Sydney, NSW 2109, Australia \\
$^{7}$Western Sydney University, Locked Bag 1797, Penrith South DC, NSW 1797, Australia \\
$^{8}$Astronomisches Rechen-Institut, Zentrum f\"ur Astronomie der Universit\"at Heidelberg, M\"onchhofstr.\ 12--14, 69120 Heidelberg, Germany \\
$^{9}$Mullard Space Science Laboratory, University College London, Holmbury St Mary, Dorking, RH5 6NT, UK \\
$^{10}$Department of Physics and Astronomy, Johns Hopkins University, 3400 N. Charles St, Baltimore, MD 21218, USA \\
$^{11}$Faculty of Mathematics and Physics, University of Ljubljana, 1000 Ljubljana, Slovenia \\
$^{12}$Observatoire astronomique de Strasbourg, Universit\'{e} de Strasbourg, CNRS, UMR 7550, 11 rue de l'Universit\'{e}, F-67000 Strasbourg, France \\
$^{13}$Research School of Astronomy \& Astrophysics, Australian National University, Cotter Rd., Weston, ACT 2611, Australia \\
$^{14}$E.A. Milne Centre for Astrophysics, University of Hull, Hull, HU6 7RX, United Kingdom \\
$^{15}$Institute of Astronomy, Madingley Rd, Cambridge CB3 0HA, UK \\
$^{16}$Kapteyn Astronomical Institute P.O. Box 800, 9700 AV Groningen, The Netherlands \\
$^{17}$INAF Astronomical Observatory of Padova, I-36012 Asiago (VI), Italy \\
$^{18}$Senior CIfAR Fellow. Department of Physics and Astronomy, University of Victoria, Victoria, BC V8P 5C2, Canada\\
$^{19}$Department of Physics, Chong Yuet Ming Physics Building, The University of Hong Kong, Hong Kong\\
}

\date{Accepted XXX. Received YYY; in original form ZZZ}

\pubyear{2017}

\begin{document}
\label{firstpage}
\pagerange{\pageref{firstpage}--\pageref{lastpage}}
\maketitle

\begin{abstract}

{We characterize the selection function of RAVE using 2MASS as our underlying population, 
which we assume represents all stars which could have potentially been observed. We evaluate the 
completeness fraction as a function of position, magnitude, and color in two ways: first, on a field-
by-field basis, and second, in equal-size areas on the sky. Then, we consider the effect of the RAVE 
stellar parameter pipeline on the final resulting catalogue, which in principle limits the parameter 
space over which our selection function is valid. Our final selection function is the product of the 
completeness fraction and the selection function of the pipeline. 
We then test if the application of the selection function introduces biases in the derived parameters. To 
do this, we compare a parent mock catalogue generated using \textsc{Galaxia} with a mock-RAVE 
catalogue where the selection function of RAVE has been applied. We 
conclude that for stars brighter than I = 12, between $4000 \rm K < T_{\rm eff} < 8000 \rm K$ and $0.5 < \mathrm{log}\,g < 5.0$, RAVE is kinematically and chemically unbiased with respect to 
expectations from \textsc{Galaxia}.}
\end{abstract}




%

\section{Introduction}
In any statistical analysis it is fundamental to understand the relation between the objects for which data were 
obtained, and the underlying population from which the sample was drawn. This relation is called the selection 
function of the sample. Without this knowledge, it is difficult to accurately infer the general properties of a 
population. 

Many large-scale astronomical surveys of Milky Way stars with data releases currently or 
soon available make some effort to characterize their selection function. The explicit 
quantification of the selection function of a stellar survey has been demonstrated by  
\citet{Schoenrich09} for the Geneva-Copenhagen survey \citep[GCS;][]{Nordstroem04}, \citet{Bovy12_sl} for a 
sub-sample of the Sloan Extension for Galactic Understanding and Exploration survey \citep[SEGUE;][]{Yanny09}, \citet{Nidever14} for the APO Galactic Evolution Experiment \citep[APOGEE;][]{Majewski15}, and \citet{Stonkute16} for the Gaia-ESO survey \citep{Gilmore12}. {A number of factors such as changes to the observing strategy, limitations due to instrumentation, or including different input catalogues can all affect the final resulting catalogue, so it is crucial to consider each of these aspects when characterising the selection function.}

In this article we present a study of the selection function of the RAdial Velocity Experiment (RAVE)  
survey based on its most recent data release \citep[DR5;][]{Kunder17}, to facilitate the wider 
and more robust use of this publicly-available catalogue. This survey was among the first surveys in 
Galactic astronomy with the explicit purpose of producing a homogeneous and well-defined data set. To 
achieve this goal, the initial target selection was based purely on the apparent $I$-band magnitudes of the stars. 

Based on the simplicity of 
the selection function a number of recent studies using RAVE data, reviewed in \citet{Kordopatis15_review}, assumed the RAVE survey to be a kinematically unbiased sample to investigate 
models of our Galaxy. In particular, \citet{Sharma14} briefly addressed the selection function with 
respect to ensuring their subsample was unbiased, by mimicking the target selection of RAVE directly using 
Monte Carlo realisations of their Galaxy models. However, here we aim to characterize the selection function of 
all stars available in DR5.

\begin{figure}
 \includegraphics[width=0.5\textwidth]{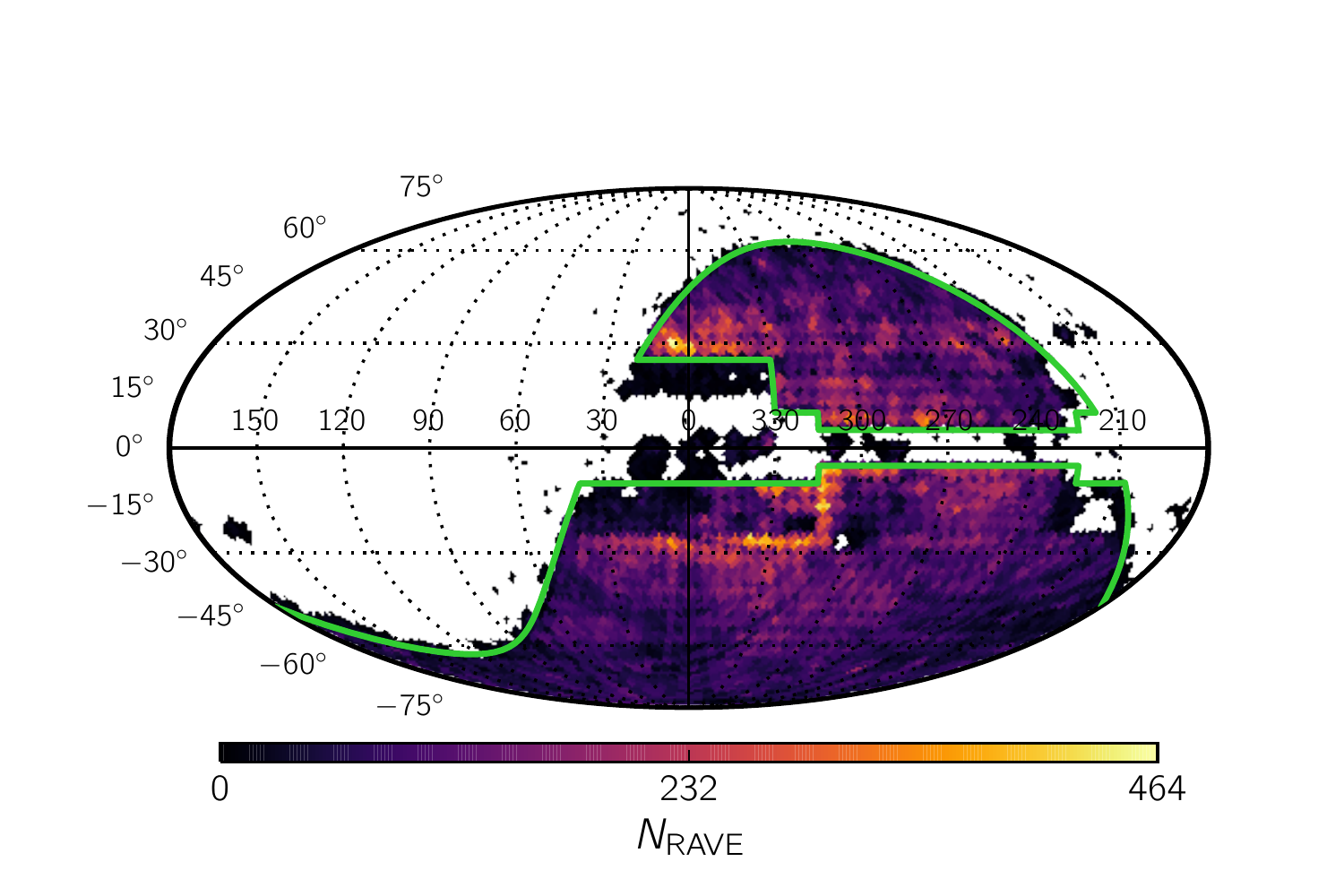}
 \caption{All RAVE DR5 targets in Galactic coordinates, colour-coded by number of stars in a given \textsc{healpix} pixel (NSIDE = 32, {area~$\simeq3.31\deg^2$}, see also Sec.~\ref{sec:healpix}). The adopted footprint (described in Sec.~\ref{sec:footprint}) is shown in green. In this study we only consider stars within the footprint.}
 \label{fig:RAVE_footprint}
\end{figure}

We present a short overview on the RAVE survey 
in Section~\ref{sec:RAVE}, summarising the history of the survey with respect to the target selection and 
observing strategy. Our reduced sample for evaluating the selection function is described in Section~\ref{sec:catalog_and_quality}. In Section~\ref{sec:selection_function} we present our results for two different 
ways of evaluating the selection function: field-by-field and by \textsc{healpix} pixel. Then in Section~\ref{sec:pipeline_impact} we incorporate the effects of the spectral 
analysis pipeline on the final catalogue. In Section~\ref{sec:Galaxia}, we 
present the method for generating our mock-RAVE catalogue, and compare it to a sample of RAVE DR5 
stars. We then test for biases due to the selection function of RAVE, by comparing our mock-RAVE catalogue 
with a parent \textsc{Galaxia} sample. Finally, we discuss the implications of these findings and our conclusions in Section~\ref{sec:conclusion}.

\section{The RAVE survey} \label{sec:RAVE}

RAVE is a large-scale spectroscopic stellar survey of the Southern hemisphere conducted using the 6dF multi-object spectrograph on the 1.2-m UK Schmidt Telescope at the Siding Spring 
Observatory in Australia, and completed in 2013. A general description of the project can be found in the data 
release papers (\citealp[DR1;][]{Steinmetz06}; \citealp[DR2;][]{Zwitter08}; \citealp[DR3;][]{Siebert11_dr3}; \citealp[DR4;][]{Kordopatis13}) as well as in the most recent data release paper \citep[DR5;][]{Kunder17}. We show the distribution of targets available in RAVE DR5 in Figure~\ref{fig:RAVE_footprint}.

The spectra were taken in the Ca\,\textsc{ii}-triplet region (8410~--~8795~\AA) with an effective spectral 
resolution of $R \approx 7500$. The strong calcium absorption lines allow a robust determination of the line-of-
sight 
velocities via the Doppler effect even with low signal-to-noise ratio (SNR) ($\lesssim$~10 per pixel). This region 
was explicitly chosen to 
coincide with the spectral range of Gaia's Radial Velocity Spectrometer (RVS) \citep{Prusti12,Bailer-Jones13,Recio-Blanco16}. While Gaia will release radial velocity and stellar parameters in forthcoming data 
releases, at present Gaia offers only position and magnitude information for approximately a billion stars \citep{Gaia16}. 
The Tycho-Gaia astrometric solution \citep[TGAS;][]{Michalik15} provides parallax and proper motion data for $\sim$ 
2 million stars which were observed by Tycho-2 \citep{Hog00}. As RAVE contains 215\,590 unique TGAS stars, 
it offers a unique advantage of providing stellar parameters for stars with improved parallax and proper motion 
data from TGAS.

\subsection{Input catalogue}
\label{sec:input_catalogue}
When observations for the RAVE survey started in 2003 there was no comprehensive photometric 
infrared survey available to serve as an input catalogue. Instead, approximate $I$-band magnitudes were calculated 
from the Tycho-2 catalogue and the SuperCOSMOS Sky Survey \citep[SSS;][]{Hambly01}, and used to 
construct an initial input catalogue of $\sim$300\,000 stars. In May 2005, the DENIS catalogue \citep{Epchtein99} became available which provided Gunn $I$-band photometry, however, it did not provide 
sufficient sky coverage to serve as the sole basis for the input catalogue. RAVE DR1, DR2, and DR3 were 
sourced from the original input catalogue \citep{Kordopatis13}. 

The fourth data release, DR4 \citep{Kordopatis13}, incorporated observations drawn from a new input 
catalogue, using DENIS DR3 \citep{DENIS05} as the basis, which had been cross-matched with the 
2MASS point source catalogue \citep{Skrutskie06}. The new input catalogue also extended the 
RAVE footprint to include lower Galactic latitudes ($5^\circ < |b| < 25^\circ$), where a colour cut using 
2MASS photometry ($J - K > 0.5$ mag) was applied to preferentially select giants \citep{Kordopatis13}. This input catalogue is also used for the most recent data release \citep[DR5;][]{Kunder17}.

\subsection{Target selection and observing strategy}
\label{sec:target_selection}

\begin{figure}
 \includegraphics[width=0.47\textwidth]{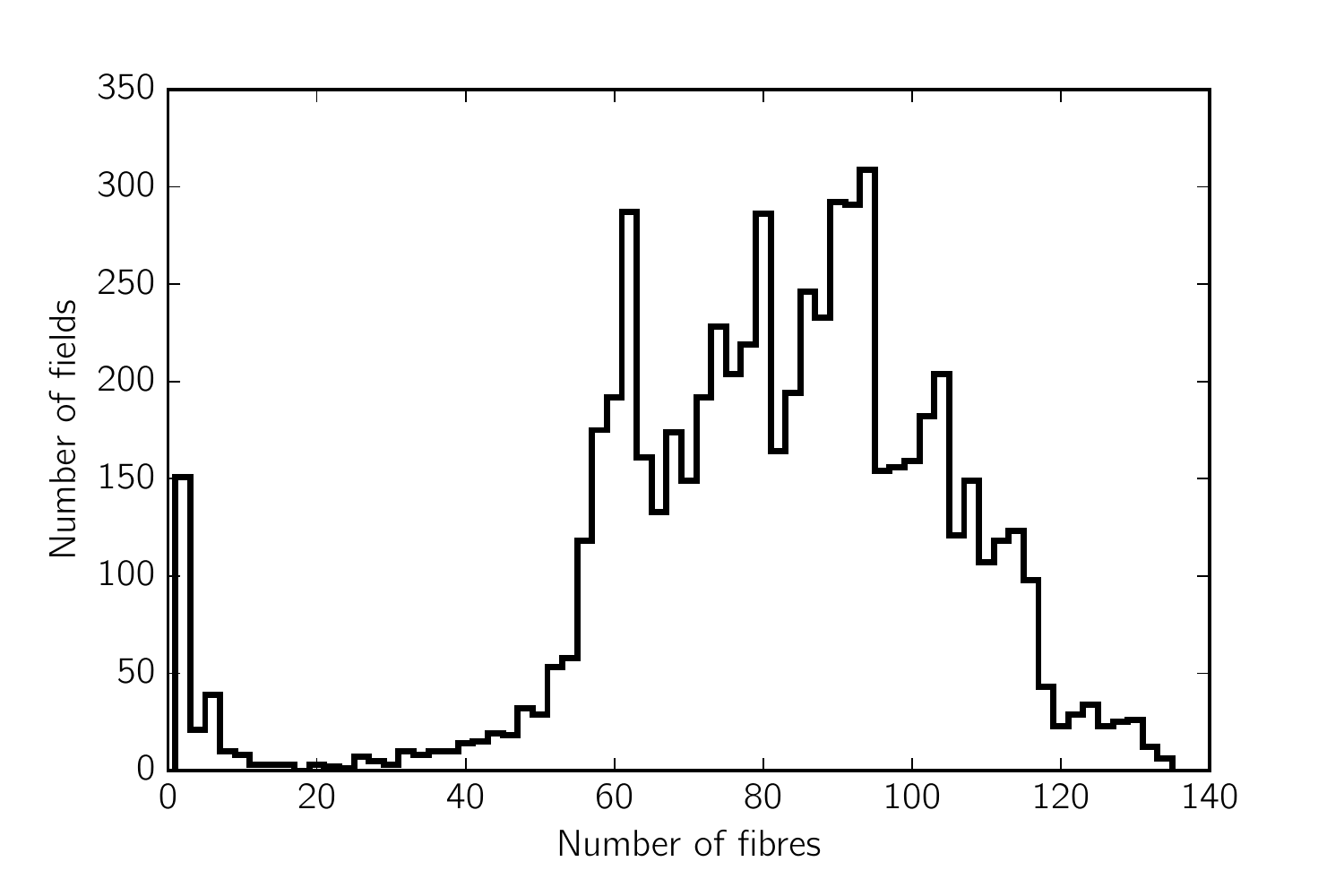}
 \caption{Histogram of the number of fibres placed for each pointing {(bin width = 2)}, for the entire duration of the RAVE survey (2003-2013). {For each pointing, at least one fibre was placed.}}
 \label{fig:fiber_hist}
\end{figure}

Here we summarise the target selection and observing strategy described in the first data release \citep[DR1,][]{Steinmetz06}, {as the selection function of a survey depends explicitly on how the observations are conducted}.

From the input catalogue described in the previous section, 400 targets
were selected for a given field of view. This selection was then split into two field files 
consisting of 200 stars each, to allow for two separate pointings. 
The 6dF instrument, used to conduct RAVE observations, consists of three fibre plates  
with 150 fibres each. 
These fibres were assigned to science targets according to a field configuration algorithm developed for the 
2dF spectrograph \citep{Lewis02}. However, for various reasons such as inaccessible areas on the fibre 
plate and fibre breakage, on average approximately 90 science fibres were allocated per pointing. {Each observation consisted of a minimum of 3 (average 5) exposures, which were then stacked to improve the SNR per pointing.}
Figure~\ref{fig:fiber_hist} shows the distribution of fibres placed on science targets present in DR5 for all fields in the master list of RAVE field centres (see Sec.~\ref{sec:fieldbyfield}). 

During the first year and a half of observations, no blocking filter
was used on the spectrograph, so spectra were contaminated with second order diffraction (i.e., flux from the 
$\sim 4200-4400 \mathrm{\AA}$ wavelength range entered the primary wavelength range). Therefore, in DR5 the automated stellar parameter pipeline does not give stellar parameters for observations made before 6 April 2004.

\begin{figure}
 \centering
 \includegraphics[width=0.47\textwidth]{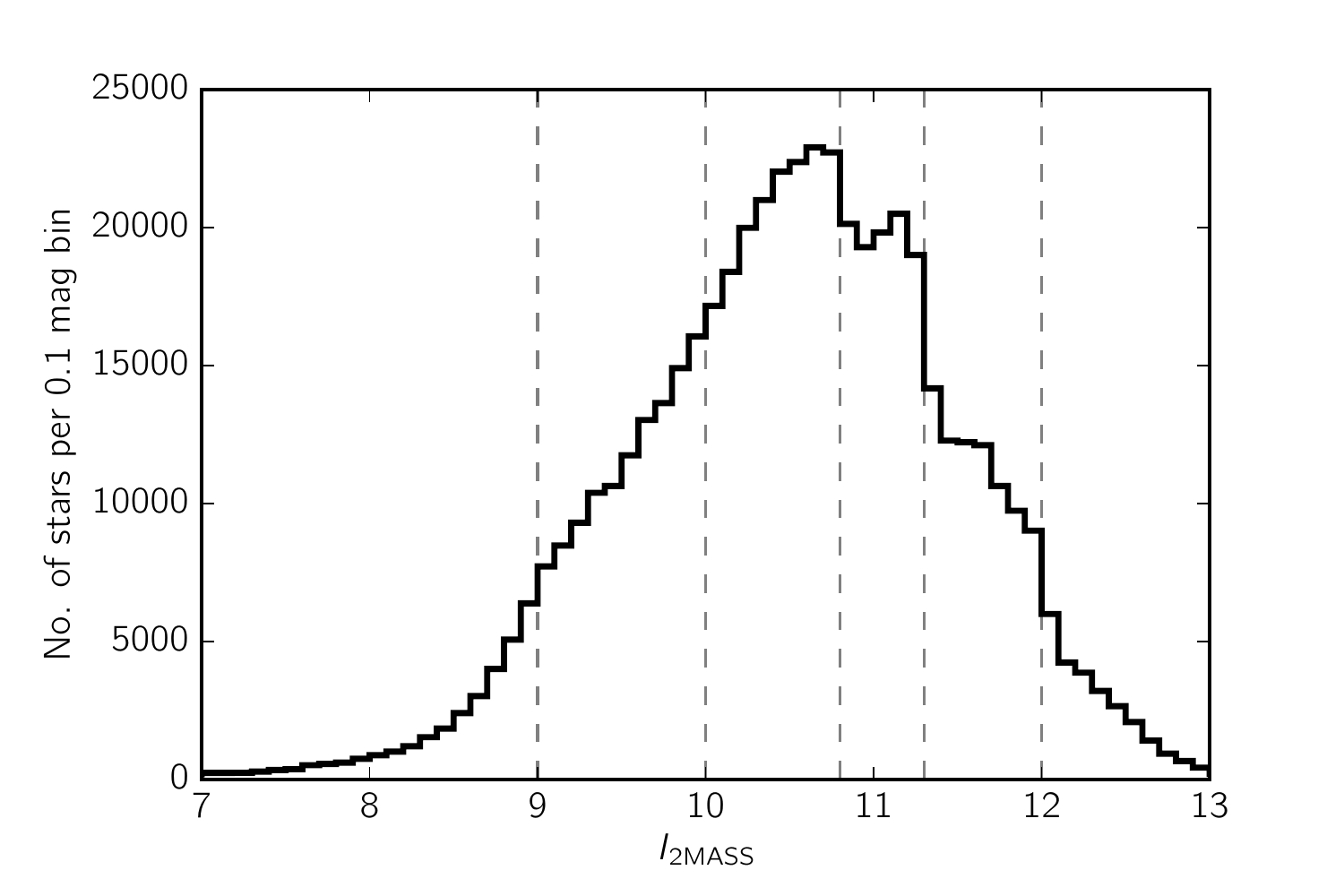}
 \caption{Distribution of $I$-magnitudes in the full RAVE DR5 catalogue. Dotted lines mark the border of the 
 magnitude bins used for observing runs after March 2006. The histogram shows $I_{\mathrm{2MASS}}$ magnitudes which are computed from the 2MASS $J$ and $K_s$ magnitudes, but present a homogeneous database.}
 \label{fig:RAVE_Imag}
\end{figure}

A problem with fibre cross-talk due to bright ($I \sim 9$) stars adjacent to fainter stars was 
also identified in the period before DR1, and corrected for in the first iteration of the data 
reduction pipeline \citep{Steinmetz06}. 
Therefore, in March 2006, the observing strategy was modified to observe stars only in a given magnitude
bin for each pointing. These magnitude bins are illustrated in Figure~\ref{fig:RAVE_Imag} as vertical dashed lines. In addition to 
reduced fibre cross-talk, this change in the 
observing strategy had the added benefit of optimizing exposure times {(e.g. bright fields could be 
observed in nominal conditions, while faint fields were preferentially observed when conditions were excellent)}, increasing the SNR per spectrum, and 
therefore resulting in more accurate stellar parameters. {For fields in which interlopers or stars with 
variable brightness affected the fibres despite the magnitude selection, assessment and data reduction 
was conducted on a case-by-case basis to minimise the probability that problematic stars would enter the final catalogue.} 

\subsection{Survey footprint}
\label{sec:footprint}

A simple footprint was imposed for observations: pointings were restricted to the Southern hemisphere and $|b| > 25^\circ$. RAVE generally avoided regions on the sky with 
large extinction, i.e., close to the Galactic disc and towards the bulge. The primary reason for avoiding low Galactic latitudes was to prevent multiple stars entering a fibre, which had a spatial extent of $7''$ on the sky. Exceptions were a number of calibration fields around $|b|=0^\circ$ and several targeted observations of 
open clusters in the Galactic plane. In addition, there are a few fields in regions at the northern side of the bulge  that originate from an interim input catalogue. We exclude these fields when 
evaluating the completeness of RAVE, as the target 
selection in these fields differed from the general selection procedure. 

In addition, we note the impact of utilizing DENIS DR3 as an input catalogue. The DENIS survey was observed in
strips of $30^\circ$ in declination and 12 arcmin in right ascension, with an overlap of 2 
arcmin between 
consecutive strips. This observing pattern is embedded in the formulation of the selection function as a function 
of position (Eq.~\ref{eq:selection_function}), and therefore is considered when evaluating both the completeness 
and selection function.

Figure~\ref{fig:RAVE_footprint} shows the adopted survey footprint for this study, which differs from the original 
footprint used for observations, as well as the 
distribution of individual stars in DR5.

\subsection{RAVE Data Release 5}
The latest public data release, DR5, contains information from 520\,781 measurements of 457\,588 individual 
stars. The distribution on the sky of these stars can be found in Figure~\ref{fig:RAVE_footprint}. In addition to 
obtaining precise line-of-sight velocities $\vlos$ (typical uncertainties $\sim2\kms$), RAVE DR5 provides 
several other stellar parameters derived from the spectra: effective temperature (\teff), 
surface gravity (\logg), an overall metallicity ([M/H]), and individual abundances for six elements: magnesium, 
aluminium, silicon, titanium, iron, and nickel.

Line-of-sight distances for RAVE stars have been estimated using a number of methods, including red-clump 
giants \citep[e.g.][]{Siebert08, Veltz08, Williams13}, isochrone fitting \citep[e.g.][]{Zwitter10, Breddels10}, and a 
robust Bayesian analysis method described in \citet{Burnett10}. RAVE DR5 provides distances derived using 
the method described in \citet{Binney14}, where stellar parameters, along with known 
positions, are used to derive spectrophotometric distance estimates for a large fraction of the stars in the 
survey. 

In addition, \citet{Matijevic12} performed a morphological classification of the spectra to allow for the 
identification of spectroscopic binaries and other peculiar stars in the catalogue. All targets in DR5 were also 
cross-matched with a number of other data sets: Tycho-2 \citep{Hog00}, UCAC4 \citep{Zacharias13}, PPMXL \citep{Roeser10}, 2MASS \citep{Cutri03}, WISE \citep{Wright10}, APASS \citep{Munari14}, and Gaia DR1 \citep{Gaia16} to provide 
additional information such as proper motions, as well as apparent magnitudes in other filter passbands. 

\section{Catalogue description and quality flags}
\label{sec:catalog_and_quality}

The RAVE survey was designed to have as simple a selection function as possible, to ensure that any 
biases could be accurately quantified. The initial target selection was based only on the apparent $I$-band 
magnitude ($9 \lesssim I \lesssim 12$) and sky position. An $I$-band selection was chosen as the most 
appropriate for efficient use of the spectral range of the 6dF instrument. In Figure~\ref{fig:RAVE_Imag} we 
show the distribution of $I$-band magnitudes in RAVE DR5. This distribution extends past the initial apparent 
magnitude limits due to {uncertainties} in the SSS photometry used for the first input catalogue \citep[see Figure 4 of][]{Steinmetz06}. 
{During 2006}, the angular footprint was expanded to include regions close to the Galactic disc 
and bulge (Galactic latitude $5^\circ < |b|<25^\circ$) {as a result of the new input catalogue (see Section~\ref{sec:input_catalogue})}, and in these new regions a colour criterion ($J-K_s \ge 0.5$) was 
imposed to select for cool giant stars over 
more prevalent dwarfs \citep{Kordopatis13}. We can thus assume that the probability, $S$, of a star being 
observed by the RAVE survey is
\begin{equation} 
\label{eq:selection_function}
 S \propto S_{\rm select}(\alpha,\delta,I,J-K_s),
\end{equation}
with $\alpha$ and $\delta$ denoting {the equatorial coordinates of stars in a given region on the sky, within the defined footprint (see Figure~\ref{fig:RAVE_footprint}).} 

Due to its complex history, and owing to observational constraints and actual atmospheric conditions on the 
respective day, the input catalogue for RAVE carries some inhomogeneity, and it is therefore not straightforward 
to construct a valid parent sample from this variety of data sets. However, one data set in particular, 
2MASS, offers complete coverage of both the survey area and the magnitude range of RAVE. Therefore, we 
adopt the 2MASS photometry in order to compare our RAVE targets with as homogenous a sample as 
possible. 

2MASS provides accurate $J$, $H$ and $K_s$ photometry for nearly all RAVE 
targets and, equally important, also for all other stars which could have potentially entered the input 
catalogue. 
Unfortunately, 2MASS does not provide $I$-band photometry, which is needed to construct our selection 
function (Eq.~\ref{eq:selection_function})\footnote{Recently, data from the APASS 
survey \citep{Munari14} became available which provides SDSS $i$ magnitudes, but this survey also 
suffers some saturation problems for bright stars. Currently, APASS is being extended to brighter magnitudes, 
so in the future this could be a valuable alternative to 2MASS.}, but we can compute an approximate $I_{\mathrm{2MASS}}$ 
magnitude via the following formula:
\begin{equation} \label{eq:I_2MASS}
  I_{\mathrm{2MASS}}- J = (J-K_s) + 0.2\exp\frac{(J-K_s)-1.2}{0.2} + 0.12
\end{equation}
{Eq.~\ref{eq:I_2MASS} is derived by a direct comparison of 2MASS $J$ and $K_s$ magnitudes with DENIS $I$ magnitudes. This transformation is determined by a polynomial fit in $I-J$ versus $J-K_s$, and is an evolution of Eq. 24 in \citet{Zwitter08}, with an improved fit for very cool stars.} The distribution of $I_{\mathrm{2MASS}}$ magnitudes for RAVE DR5 is shown in Figure~\ref{fig:RAVE_Imag}. 
Here, we find a significant number of RAVE stars which have $I_{\mathrm{2MASS}} < 9$. We note that this is 
due to the fact that both DENIS and SuperCOSMOS saturate around $I_{\mathrm{DENIS}} \sim 9$, and the 
conversion of their cross-matched 2MASS magnitudes gives magnitudes brighter than $I_{\mathrm{2MASS}} \sim 9$.

In addition, there are a number of
other factors which also have an influence on the final selection function, which we will describe in the following 
sections.

\subsection{Sample selection}

\begin{table}

 \centering
 \caption{Quality criteria for the 2MASS parent sample.}

 \begin{tabular}{|l|c|l|}
  \hline
  Criterion & Requirement & Description \\
  \hline\hline
  \texttt{ph\_flag}$_{J}$ & A,B,C or D & good $J$-photometry\\
  \texttt{ph\_flag}$_{K_s}$ & A,B,C or D & good $K_s$-photometry\\
  \texttt{cc\_flag}$_{J}$ & 0 & no artifact/confusion\\
  \texttt{cc\_flag}$_{K_s}$ & 0 & no artifact/confusion\\
  \texttt{gal\_contam} & 0 & not contaminated by\\
                       &   & extended source\\
  \texttt{pm\_flag} & 0 & not associated with \\
                    &   & asteroid/comet\\
  \hline
 \end{tabular}
 \label{tab:2MASS_qc}
\end{table}

\begin{table*}
 \caption{Completeness fraction of RAVE on a field-by-field basis, for 0.1\,mag width bins.}
 \begin{tabular}{l l l l l l l l l l l l l}
  \hline 
  & & & & & & \multicolumn{7}{|c|}{Completeness Fraction $\left(I_{\rm 2MASS}\right)$} \\
  RAVE Field & Field Center & Observation Date & Fibres Placed & Spectra & 2MASS & 0.0 & ... & 9.9 & 10.0 & 10.1 & ... & 14.0\\
  Index & $(\alpha,\delta)$ & YYYYMMDD & N & N & N & & & & & & &\\
  \hline \hline
  0 & (143.20, -8.75) & 20030412 & 124 & 121 & 14170 & - & & 0.0 & 0.0 & 0.0 & & 0.00017\\
    \vdots & & & & & & & & & & & &\\
  1077 & (181.67, -7.27) & 20060418 & 93 & 93 & 7915 & - & & 0.333 & 0.391 & 0.273 & & 0.0\\ 
  1078 & (209.60, -26.43) & 20060418 & 96 & 96 & 16382 & - & & 0.167 & 0.156 & 0.156 & & 0.0\\ 
  \vdots & & & & & & & & & & & &\\
  6430 & (222.48, -44.90) & 20130404 & 86 & 85 & 69564 & - &  & 0.0 & 0.0 & 0.0 & & 0.0\\
  \hline \\ 
 \end{tabular}
 \label{tab:field_by_field}
\end{table*}

\begin{table*}
 \caption{{Completeness fraction of RAVE on a pixel-by-pixel basis, for 0.1\,mag width bins. Here, the nested scheme is used to determine a given pixel ID.}}
 \begin{tabular}{l l l l l l l l}
  \hline 
  & & \multicolumn{6}{|c|}{Completeness Fraction $\left(I_{\rm 2MASS}\right)$} \\
  \textsc{healpix} Pixel ID & 0.0 & ... & 9.9 & 10.0 & 10.1 & ... & 14.0\\
  Index (Nested) &  & & & & & &\\
  \hline \hline
  0 & - & & 0.200 & 0.667 & 0.600 &  & 0.0\\
    \vdots & & & & & & &\\
  10000 & - & & 0.250 & 0.462 & 0.500 &  & 0.0\\ 
  10001 & - & & 0.571 & 0.500 & 0.250 &  & 0.0\\ 
  \vdots & & & & & & & \\
  12287 & - &  & 0.333 & 0.333 & 0.600 &  & 0.0\\
  \hline \\ 
 \end{tabular}
 \label{tab:pix_by_pix}
\end{table*}

\subsubsection{RAVE quality criteria} 
\label{sec:quality_rave}

{To asses the completeness $S_{\mathrm{select}}$ 
(Eq.~\ref{eq:completeness}), we remove fields which were reprocessed during the course of data 
reduction (indicated in DR5 with either `a',`b', or `c' appended to the {\tt RAVE\_OBS\_ID}). After 
removing these stars, we are left with a sample of 518\,079 entries in DR5, corresponding to 455\,626 
individual spectra.}

\subsubsection{2MASS quality criteria}
\label{sec:quality_2mass}

We compute an $I_{\rm 2MASS}$ value (Eq.~\ref{eq:I_2MASS}) for each 2MASS star and clean the data 
from spurious measurements. Our requirements for a {`valid'} measurement are given in Table~\ref{tab:2MASS_qc}. 

\section{The selection function} \label{sec:selection_function}
\subsection{Field-by-field} \label{sec:fieldbyfield}

We first consider the selection function of RAVE on a field-by-field basis, in order to account for 
changes in the observing strategy as a function of time. 

First, the observation date and position for each individual pointing is identified from a master list of RAVE field 
centres and their corresponding given {\tt RAVE\_OBS\_ID}.  In order to make the 
most accurate comparison between RAVE and our parent 2MASS sample, we must utilize accurate field centre positions.
We identify 6593 individual pointings from this master list, corresponding to 1598 unique field centres. Next, we compare this list with a table containing 
information about the placement of fibres for each pointing. For each pointing, we count the number of 
fibres placed on science targets, as well as how many fibres were assigned to the sky, or simply not used. As 
shown in Figure~\ref{fig:fiber_hist}, out of the 150 available fibres on 6dF, at maximum approximately 130 fibres 
 were placed per field pointing, with an average of approximately 90 fibres per 
 RAVE pointing. From the fibres placed on science targets, we then consider how many of these observations 
 obtained spectra for which stellar parameters are published in DR5, and characterize the $I_{\mathrm{2MASS}}$ distribution by 
 counting the number of stars per 0.1\,dex magnitude bin. Then, for each RAVE pointing we determine the number of 2MASS stars available in each magnitude bin, with the quality criteria described in Sec.~\ref{sec:quality_2mass} applied. {For the final table (Table~\ref{tab:field_by_field}), we include only those fields which have stars parametrized and published in DR5.}

\begin{figure*}
 \includegraphics[width=0.8\textwidth]{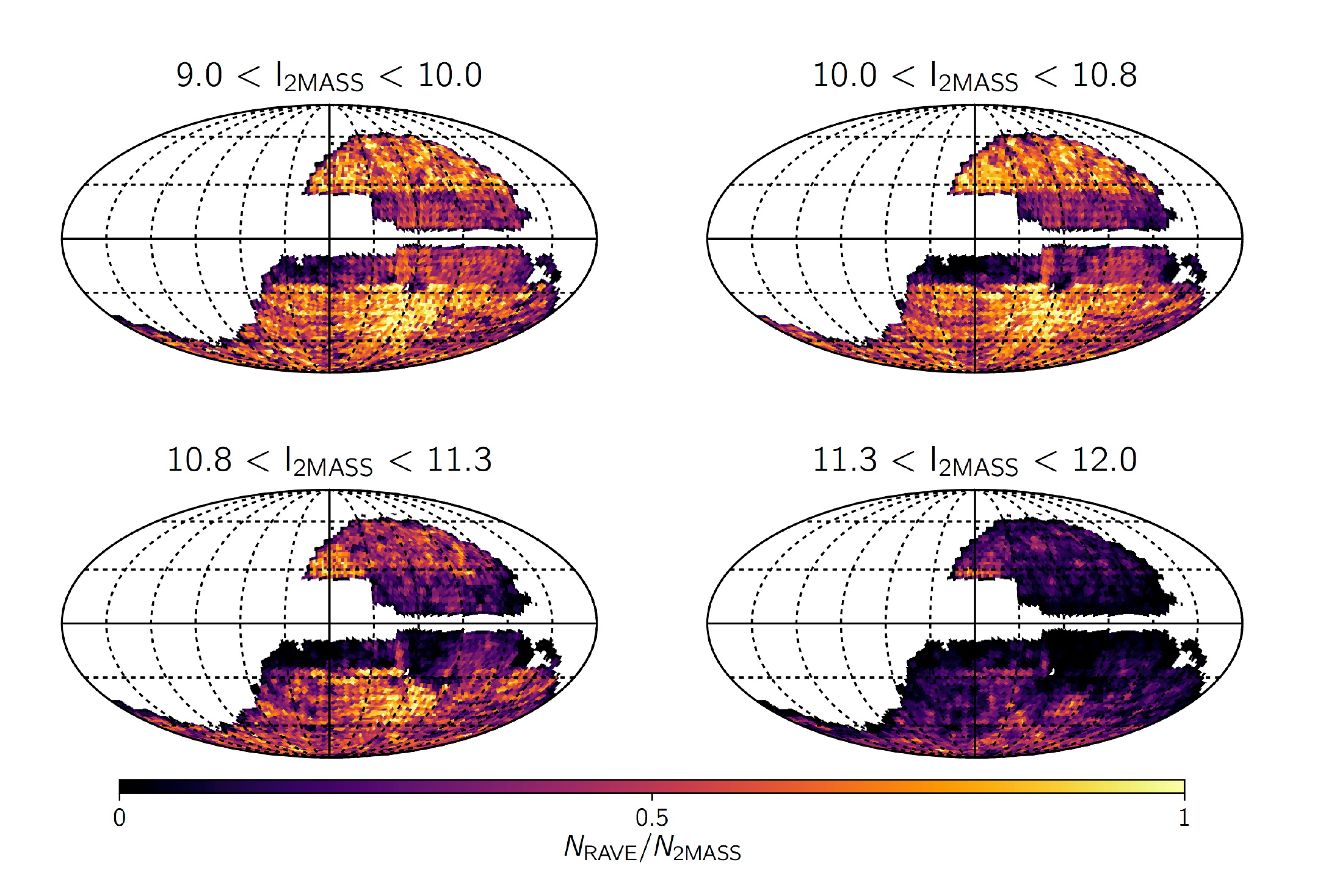}
 \caption{Completeness of RAVE DR5 in Galactic coordinates as a function of observed magnitude bins  {\citep[compare with similar plot for completeness fraction of DR4, Figure~3 of][]{Kordopatis13}}. The \textsc{healpix} pixels are colour-coded by the fractional completeness, ($N_{\mathrm{RAVE}}/N_{\rm 2MASS}$).}
 \label{fig:completeness_r_2m}
\end{figure*}

{An excerpt of the} resulting completeness fraction on a field-by-field basis can be found in Table~\ref{tab:field_by_field}. The completeness fraction for a field centered on ($\alpha, \delta$) is given by 
\begin{equation}
\label{eq:completeness_fbf}
  S_{\rm select} (\rm field_{\alpha,\delta}) =
   \frac{ \sum \sum N_{\rm RAVE}(\rm field_{\alpha,\delta},I,J-K_s)}{\sum \sum N_{\mathrm{2MASS}}(\rm field_{\alpha,\delta},I,J-K_s)} ,
\end{equation}
where the double sum is over a given $I_{\rm 2MASS}$ range and the total $J - K_s$ range in that field.

\begin{figure*}
 \includegraphics[width=\textwidth]{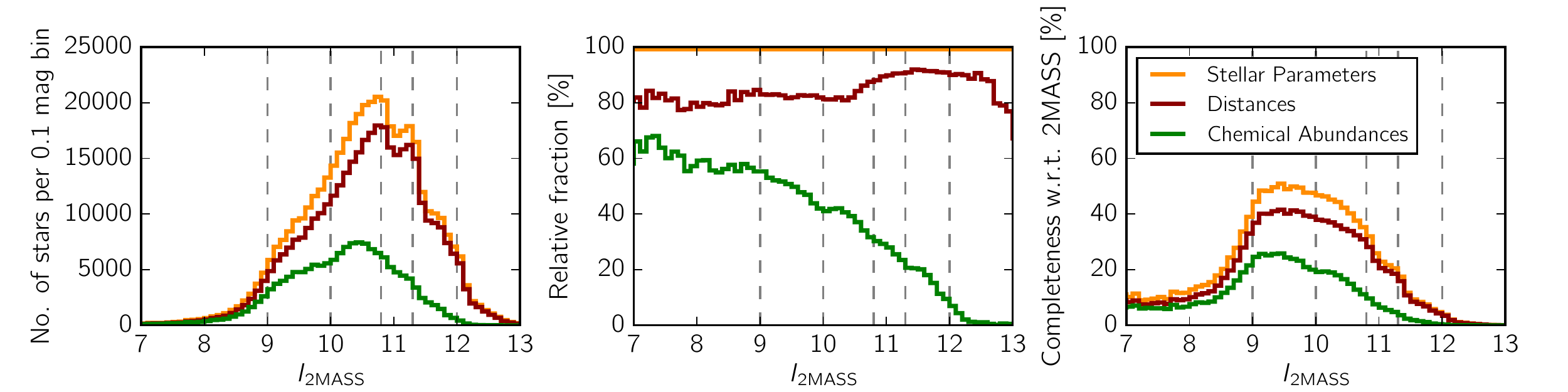}
 \caption{\textit{Left:} Histogram of stellar parameters, chemical abundances, and distance measurements in 
RAVE DR5 which satisfy the quality criteria and parameter limits given in Section~\ref{sec:pipeline_impact}, as 
a function of magnitude {($S_{\rm pipeline}$, Eq.~\ref{eq:s_pipeline})}. Stars with stellar parameters are indicated in 
orange, distances in red, and chemical abundances in green. Observed magnitude bins are indicated with 
dashed lines. \textit{Middle:} {relative fraction of stars with derived parameters as a function of magnitude. We use radial velocity as a baseline for comparison, as all stars satisfying the criteria given in Section~\ref{sec:pipeline_impact} have radial velocity measurements. As all stars with radial velocities in this sample also have stellar parameters, the completeness of stellar parameters is 100 per cent.} \textit{Right:} 
completeness fraction of derived parameters, relative to the number of 2MASS stars, {as a function of magnitude.} {This represents the 
complete selection function with respect to 2MASS (see Eq.~\ref{eq:s_total}}).}
 \label{fig:completeness_hist}
\end{figure*}

\begin{figure}
 \includegraphics[width=0.47\textwidth]{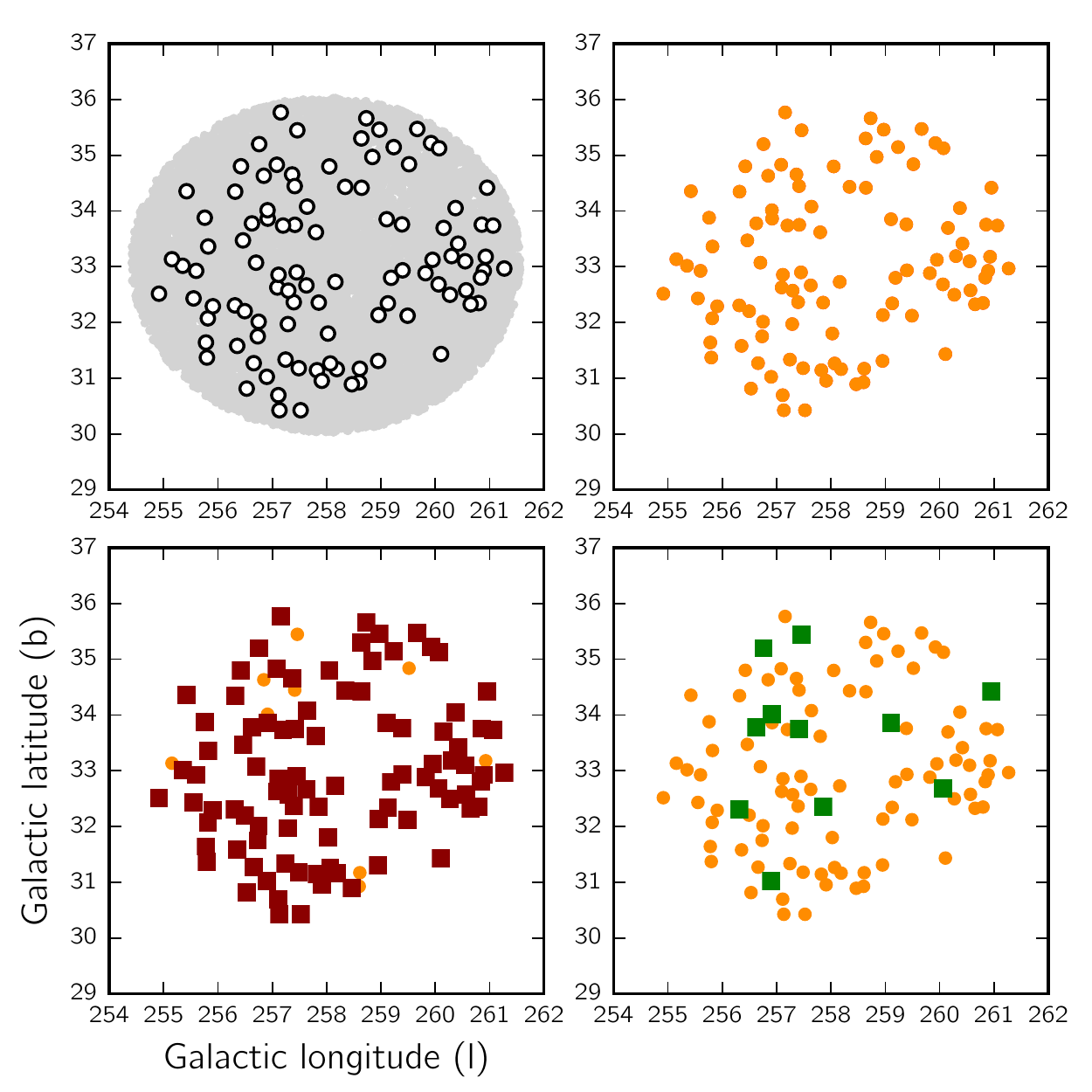}
 \caption{Distribution of RAVE stars (open circles) and 2MASS (grey) stars on the sky, for a given RAVE pointing. Orange indicates that a given RAVE star has spectral parameters from the spectral parameter pipeline ($T_{\mathrm{eff}}$, log$g$, [M/H]), red {squares} indicate stars which have distance estimates from the distance pipeline, and green {squares} indicate stars which have all abundance measurements from the chemical abundance pipeline.}
 \label{fig:completeness_rave_field}
\end{figure}

It is important to note that there exists substantial overlap between RAVE pointings, and therefore it is not appropriate to combine the data given in Table~\ref{tab:field_by_field} to construct a selection function for the 
entire RAVE survey. In order to facilitate this, we must consider the completeness of RAVE for equal, discrete 
areas on the sky. We do note however, that on scales below the size of the field plate ($\simeq 28.3\deg^2$), 
we expect inhomogeneities due to certain technical constraints with fibre positioning on the field plates used for 
RAVE observations (see Figure 3 of \citealt{Steinmetz06}).

\subsection{Equal area on the sky (\textsc{healpix})}
\label{sec:healpix}

{To construct our parent RAVE sample for considering equal areas on the sky, we first remove all 
repeat observations and keep for each star only the observation with the highest SNR. This is in contrast to Section~\ref{sec:fieldbyfield}, where we do not remove duplicates. Here, the goal is not to 
conserve the temporal information, but to accurately reconstruct the sky coverage and completeness of RAVE, 
so any given star is counted only once, even if it was observed multiple times. In addition, for the rest of the study we will only consider stars within the adopted footprint (Fig.~\ref{fig:RAVE_footprint}). This excludes $\sim7000$ stars available in RAVE DR5. These specific stars are documented in the RAVE DR5 catalogue with {\tt FootPrint\_Flag}.}

We then divide the sky into equal area pixels using the \textsc{healpix} algorithm \citep{Gorski05}. As described 
in the previous section, using the RAVE fields directly would cause additional complications for certain 
applications because some fields are overlapping. We use 12\,288 pixels for 
the whole sky ({NSIDE = 32}) which results in a pixel area of $\simeq3.36\deg^2$, much smaller than 
the size of a RAVE field ($\simeq 28.3\deg^2$). {We note that we use the `nested'\footnote{The nested, or tree structure, scheme refers to the way that \textsc{healpix} pixels are numbered \citep[see Figure 4 of][]{Gorski05}. The hierarchical structure of the nested scheme allows for degrading the resolution of a \textsc{healpix} map from the base resolution, and is the same scheme used for Gaia DR1.} scheme and equatorial coordinates ($\alpha, \delta$) to determine the corresponding pixel ID for any given star.} We count the 
number of RAVE stars, $N_{\mathrm{RAVE}}$, in each pixel (centered on $\alpha$ and $\delta$) as a function of $I_{\mathrm{2MASS}}$ in 0.1~dex 
magnitude bins. To estimate the completeness we follow the same procedure for all stars in our 2MASS sample 
to obtain $N_{\mathrm{2MASS}}$ and then compute

\begin{equation}
\label{eq:completeness}
  S_{\rm select} (\rm pixel_{\alpha,\delta}) =
    \frac{\sum \sum N_{\rm RAVE}(\rm pixel_{\alpha,\delta},I,J-K_s)}{\sum \sum N_{\rm 2MASS}(\rm pixel_{\alpha,\delta},I,J-K_s)} ,
\end{equation}
where the double sum is over a given $I_{\rm 2MASS}$ range and the total $J - K_s$ range in that pixel. {Table~\ref{tab:pix_by_pix} gives an excerpt of the completeness fraction for \textsc{healpix} pixels, in 0.1 magnitude width bins. Full versions of Tables~\ref{tab:field_by_field} and~\ref{tab:pix_by_pix} are available as part of the online-only materials, and also via the RAVE website.}

The resulting completeness as a function of 
magnitude and sky position has already been shown in the fourth RAVE data release paper \citep[Figure~3 of][]{Kordopatis13}, and we replicate it here for DR5 in Figure~\ref{fig:completeness_r_2m}\footnote{We note that the completeness fraction can, in some very rare cases, be null or greater than one. This is due to the fact that we remove stars from our parent 2MASS sample that do not meet the specified quality criteria.}.
Overall, as in DR4, we find the completeness is highly anisotropic on the 
sky for any given magnitude bin, and drops off significantly for fainter magnitudes.

\subsection{Impact of the analysis pipeline} \label{sec:pipeline_impact}
Until now we have only investigated effects that originate from the RAVE target selection. However, when 
considering certain applications, there is another important issue: namely, the effects of the automated 
pipelines. 
RAVE DR5 contains output from a number of pipelines which provide additional information for observed stars. 
As described in Sec.~\ref{sec:RAVE}, in addition to line-of-sight velocities, RAVE provides estimates of stellar 
parameters such as effective temperature, surface gravity, elemental abundances, as well as distance and age 
estimates. 

Here we consider the completeness fraction of stars with assigned stellar parameters from the stellar 
parameter pipeline, following the recommendations given by \citet{Kordopatis13}, selecting all stars that have
\begin{itemize}
 \item SNR $\ge 20$,
 \item $|${\tt correctionRV}$| < 10\kms$,
 \item $\sigma(\vlos) < 8\kms$,
 \item {\tt correlationCoeff} $>10$ (\citet{Tonry79} correlation coefficient).
\end{itemize}

In addition, this pipeline yields reliable results only in a restricted region in stellar parameter space 
\citep{Kordopatis13}. We explicitly implement this by using only stars with
\begin{equation} \label{eq:pipeline-limits}
 \begin{array}{rcccl}
  4000\rm K &<& T_{\mathrm{eff}} &<& 8000\rm K\,, \\
  0.5 &<& \mathrm{log}\,g&<& 5\,. \\
 \end{array}
\end{equation}
 
These limits are based on the range of parameters for the spectra used for the 
learning grid of the analysis pipeline \citep{Kordopatis11, Kordopatis13}, as well as unphysical or highly unlikely 
combinations of derived parameters. 

These restrictions have to be taken into account when comparing observed data with specific Galaxy models. 
They can be expressed as an additional selection function
\begin{equation}
\label{eq:s_pipeline}
S_{\mathrm{pipeline}} =  S_{\mathrm{pipeline}}(T_{\mathrm{eff}},\mathrm{log}\,g,\mathrm{[Fe/H]})
\end{equation}
and hence the complete selection function $S$ is
\begin{equation}
\label{eq:s_total}
 S = S_{\mathrm{pipeline}} \times S_{\mathrm{select}}.
\end{equation}

We give examples of this effect in Figures~\ref{fig:completeness_hist} and \ref{fig:completeness_rave_field}, for 
the selection function evaluated with \textsc{healpix} pixels, and field-by-field, respectively. 
Figure~\ref{fig:completeness_hist} shows the {distribution of the number of stars satisfying these criteria  
that have derived parameters (stellar parameters, distance, and chemical abundances) available in RAVE DR5} as a function of $I_{\mathrm{2MASS}}$ magnitude {(left and middle panels represent $S_{\rm pipeline}$, see Eq.~\ref{eq:s_pipeline})}, as well as the completeness fraction of these parameters in RAVE with respect to 2MASS {(right panel represents the complete selection function, see Eq.~\ref{eq:s_total})}. We find that the number of stars {having a given parameter in DR5} varies as a function of magnitude, with the brightest magnitude bin ($9 < I_{\mathrm{2MASS}} < 10$) having the highest number of stars with stellar parameters, distances, and chemical abundances. {When we consider the relative fraction of stars with a given parameter (using radial velocity as a baseline, as all stars satisfying the quality criteria have radial velocity measurements), we find stellar parameters} are derived for all stars with radial velocities, while distances are derived for $\sim 80$ per cent of these stars. The {relative fraction of stars with} 
chemical abundance estimates is calculated for stars which have all six element abundances 
derived from the chemical abundance pipeline \citep{Boeche11}. We find that $\sim 40-60$ per cent of stars 
brighter than 10th magnitude have chemical abundance information available in DR5. {Finally, when we consider the completeness of a given derived parameter in RAVE with respect to 2MASS, we find that stars in the brightest magnitude bin ($9 < I_{\mathrm{2MASS}} < 10$) have the highest completeness. This panel represents the complete selection function (see Eq.~\ref{eq:s_total}).}

In Figure~\ref{fig:completeness_rave_field}, we characterize the completeness fraction of derived parameters for a typical RAVE pointing. RAVE stars are shown in black, purple, and orange, with the underlying 2MASS parent sample shown in grey. For this particular pointing, we find all stars have estimated stellar parameters, $\sim 90$ per cent have distances, and $\sim 10$ per cent have chemical abundance estimates. 

\section{Comparison with a Galactic model}
\label{sec:Galaxia}

\begin{figure*}
 \includegraphics[width=0.7\textwidth]{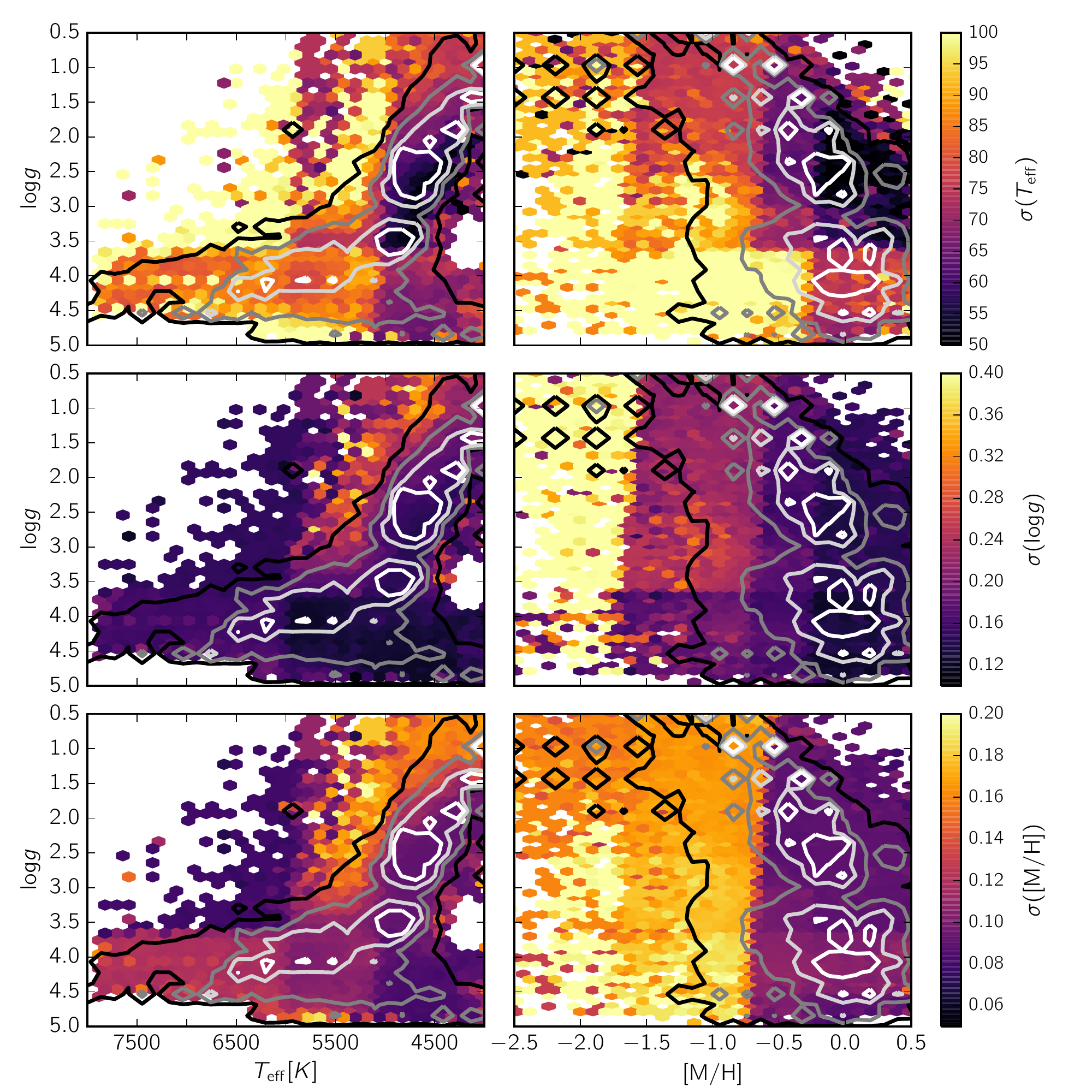}
 \caption{Mean uncertainties as a function of stellar parameters {available in RAVE DR5}. The left column shows the distribution of {uncertainties} 
 in $T_{\mathrm{eff}}$- log$g$ space, and the right column shows the same but in [M/H]-log$g$  
 space. Each row shows the distribution of the uncertainties of a different parameter as indicated by the colour 
 bars on the far right. The contours indicate 33, 67, 90, and 99 per cent of all RAVE DR5 stars.}
 \label{fig:hist_errors}
\end{figure*}

We now explore the potential influence of the selection function with respect to inducing biases in the stellar 
parameter distributions of our RAVE DR5 stars compared with what we expect from models of the Galaxy. For 
this comparison, we utilize the stellar population synthesis code \textsc{Galaxia}\footnote{\href{http://galaxia.sourceforge.net/}{http://galaxia.sourceforge.net/}} \citep{Sharma11}.

\textsc{Galaxia} is a tool which uses a given Galactic model to conduct synthetic observations, generating a 
catalogue which imitates any given survey of the Milky Way. Here, we use the default provided in \textsc{Galaxia}, a modified version of the Besan\c{c}on model \citep{Robin03}. Details on the extent of these 
modifications can be found in \citet{Sharma11}. The Besan\c{c}on 
model within the \textsc{Galaxia} framework has been found to agree quite well with Besan\c{c}on star counts \citep{Sharma11}. The input parameters for \textsc{Galaxia} are very simple, and 
correspond well to our adopted form of RAVE's selection function (Eq.~\ref{eq:selection_function}). 

The catalogue may be generated for a given circular area on the sky, 
as well as for the whole sky. In order to compare these mock observations with our two methods of 
characterising the selection function of RAVE, we generate two catalogues: one on a field-by-field basis, and 
one full-sky, which is then divided into \textsc{healpix} pixels. For each of these catalogues, we allow 
\textsc{Galaxia} to generate stars with apparent $I$-band magnitude $0 < I < 13$, and no colour restriction. {We then perturb the output from \textsc{Galaxia} with a simple noise model to imitate 
observational uncertainties present in RAVE, and apply the RAVE selection function. We refer to this modified 
catalogue as our `mock-RAVE' catalogue. The mock-RAVE catalogue can then be compared to our parent 
Galaxia sample (where the RAVE selection function has not been applied), to evaluate the effect that the selection function has on fundamental distributions such as 
kinematics and chemistry.}

\begin{figure*}
\begin{centering}
 \includegraphics[width=0.9\textwidth]{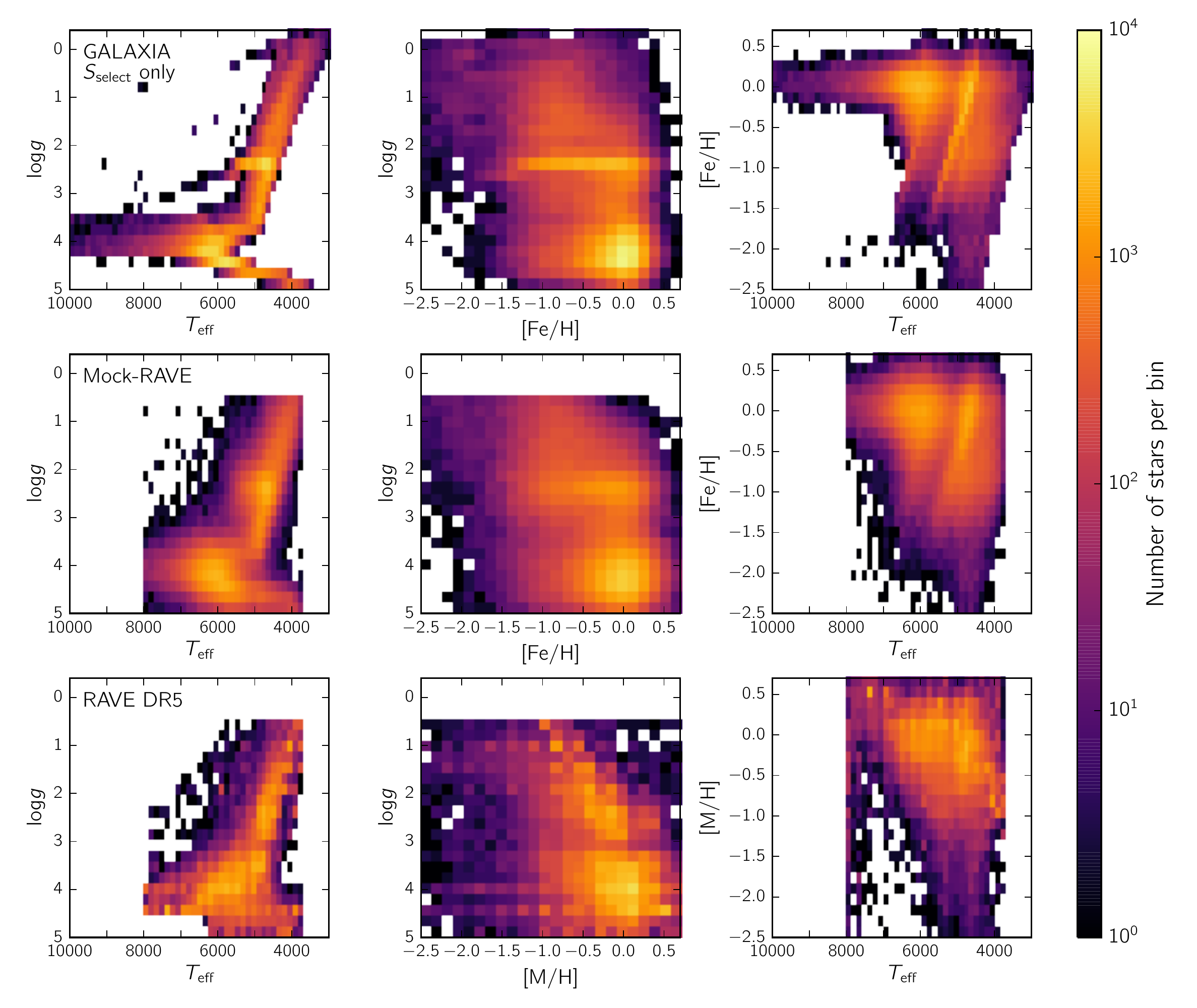}
 \caption{2D histograms of three stellar parameter spaces:  log$g$-$T_{\mathrm{eff}}$ (left), log$g$-abundance ([Fe/H] for \textsc{Galaxia}, [M/H] for RAVE) (middle), and abundance-$T_{\mathrm{eff}}$(right). The top row shows these 2D histograms for our \textsc{Galaxia} sample with the RAVE selection function applied. In the middle row we show our \textsc{Galaxia} sample which has had both the RAVE selection function and RAVE-like {uncertainties} applied. The bottom row shows our RAVE DR5 sample. The colour scale is log normalized.}
 \label{fig:2dhist_compare}
 \end{centering}
\end{figure*}

\subsection{Applying uncertainties to generate a mock-RAVE catalogue}
\textsc{Galaxia} provides stellar parameters and magnitudes with infinite precision and accuracy. This does not 
reflect our observational data, where each of the derived parameters has intrinsic {uncertainties associated with} its 
measurements. {In order to facilitate an accurate comparison between the mock catalogue and real RAVE 
data, we perturb $J$, $K_s$, $T_{\rm eff}$, log $g$, and 
[Fe/H] available in our \textsc{Galaxia} catalogue based on the uncertainty distributions of 2MASS magnitudes and RAVE stellar 
parameters before applying the selection function of RAVE. We then apply the selection function of RAVE using both methods described in Sec.~\ref{sec:selection_function}: field-by-field and \textsc{healpix} pixels. In addition to scattering the \textsc{Galaxia} distributions with our simple noise model, we slightly 
modify the metallicity distribution of the thick disc and the halo of our \textsc{Galaxia} output, for better 
agreement with observations.}

\subsubsection{2MASS apparent magnitude uncertainties}
{First, we modify the output \textsc{Galaxia} 2MASS $J$ and $K_s$ magnitudes by a simple noise model, derived from the observational uncertainties in 2MASS.} To do 
this, we characterize the {observational uncertainty} for a given 0.1 magnitude bin as a function of magnitude. We model the 
distribution of {uncertainties} in each bin as a Gaussian, and draw from this Gaussian to obtain an `observational {uncertainty'} on our \textsc{Galaxia} output. Typical 2MASS $J$ magnitude {uncertainties} are of the order of 0.025 dex. From the modified $J$ and $K_s$ values, we obtain an $I_{\mathrm{2MASS}}$ for each \textsc{Galaxia} star using Eq.~\ref{eq:I_2MASS}. 

\subsubsection{Applying RAVE-like uncertainties to stellar parameters}

In order to compare the stellar parameters available in this mock catalogue with those derived from the RAVE 
DR5 stellar parameter pipeline, we must first modify the output from \textsc{Galaxia} with the {uncertainty}  distributions 
of RAVE stellar parameters. The RAVE DR5 stellar parameter pipeline provides individual uncertainties for 
each star, and we can use the distribution of these uncertainties to modify our initial \textsc{Galaxia} catalogue 
by {RAVE-like uncertainties}, similar 
to the process used in the previous section, but in a higher-dimensional space due to correlations between the 
uncertainties. 

In Figure~\ref{fig:hist_errors}, we show the correlation of uncertainties as a function of position in different 
planes of stellar parameters. Here, we colour-code the mean uncertainty as a function of the stellar 
parameters in $T_{\mathrm{eff}}$- log$ g$ and $T_{\mathrm{eff}}$-[M/H] space. The highest uncertainties are found primarily in hot, giant stars in the $T_{\mathrm{eff}}$- log$ g$ plane, and metal-poor stars in 
the [M/H]-log$g$ plane (see also Table 4, \citealt{Kunder17}). However, comparing these regions to the density contours, we find 
that these regions are sparsely populated, and therefore should not significantly affect the mean {uncertainty}. The abrupt jumps, visible at e.g. $T_{\rm eff} \sim 5000\,$K and [M/H] $\sim -0.7$, result from discrete coverage of the stellar parameter space by model atmospheres that are compared to the observed spectra by the pipeline. We find that the majority of RAVE stars have similar {uncertainties} in spectral parameters, with $\langle \sigma(T_{\mathrm{eff}}) 
\rangle \sim 50-75\, \mathrm{K}$, $\langle \sigma(\mathrm{log} g) \rangle \sim 0.1-0.2$ dex, and $\langle 
\sigma(\mathrm{[M/H]}) \rangle \sim 0.1$ dex. 

In addition to an anisotropic distribution of {uncertainties} in in $T_{\mathrm{eff}}$- log$ g$ and $T_{\mathrm{eff}}$-[M/H] 
space, it has been well documented that these {uncertainties} in the derived atmospheric parameters are also correlated (see Figure 6 of \citet{Kordopatis11} and Figure 23 of \citealt{Kordopatis13}). 
Due to these correlations, it is not sufficient to simply model the {uncertainties} as individual Gaussians 
and draw from them. Instead, we consider the distribution of {uncertainties} to have the form of a multivariate Gaussian, 
and estimate the covariance between {uncertainties} in $T_{\mathrm{eff}}, \mathrm{log} g$, and [M/H]. We 
then draw from this multivariate Gaussian to obtain simultaneously {uncertainties} for these three 
respective parameters. Note that in this way we can introduce only the internal uncertainties of the analysis 
pipeline, but not systematic shifts coming from inaccuracies of the stellar atmosphere models.

Finally, we apply $S_{\rm pipeline}$ by setting weights to zero for all stars that do not fulfill the criteria given in Eq.~\ref{eq:pipeline-limits}. {We refer to the result as the mock-RAVE 
catalogue.} The effect of this step is model dependent as, for example, the number of 
super-solar metallicity stars varies between different Galaxy models. Using the version of the Besan\c{c}on 
model in \textsc{Galaxia}, we find that approximately 9 per cent of stars fall outside of our $T_{\mathrm{eff}}$ and $\mathrm{log}g$ limits.

The {effect of applying these observational uncertainties as well as $S_{\rm pipeline}$} is shown in Figure~\ref{fig:2dhist_compare}. The top row shows 2D 
histograms of stellar parameters for our \textsc{Galaxia} sample (without the application of $S_{\rm pipeline}$). RAVE-like {uncertainties} and the selection function 
are applied to obtain the panels in the middle row. Our RAVE sample is shown on the bottom row. Overall, we 
find good agreement in the distribution of these stellar parameters between the observations and the mock-RAVE catalogue.

\subsection{Impact of the selection function} \label{sec:Impact}
We now turn to the implications of the observed stellar populations due to the selection function of RAVE. While 
RAVE targets within the footprint  were selected on purely photometric grounds, it remains to be seen if changes to the observing 
strategy as well as the applied colour cut at low latitudes have induced biases in the observed characteristics of 
the sample. In order to test if RAVE is a kinematically unbiased survey, we compare the Galactocentric 
cylindrical velocity distributions of the parent \textsc{Galaxia} sample with those of the mock-RAVE catalogue. We also 
examine potential biases in the metallicity distribution of the sample, as abundance measurements are 
highly correlated with other derived values, such as effective temperature and surface gravity, as well as 
external characteristics such as kinematics. Hence, biases in either velocity or metallicity are potentially 
harmful if undetected, for both chemical evolution and dynamical modeling. 

We take a uniformly selected subsample of our full \textsc{Galaxia} catalogue in the footprint of RAVE as our expected 
`parent' sample (i.e., what we consider to be the `truth' for the purpose of this exercise), and 
compare it to our mock-RAVE catalogue. Any considerable deviations between the two 
distributions may indicate a bias in RAVE due to the selection function. We note that for this 
exercise, we do not apply RAVE-like {uncertainties} to the velocities or metallicities in our mock-RAVE 
catalogue (i.e., here we use the true \textsc{Galaxia} output). In addition to a \textsc{Galaxia} subsample limited to $I < 13$, we 
also investigate the effects of limiting our \textsc{Galaxia} subsample to $I < 12$, as it has been shown in Figure~\ref{fig:RAVE_Imag} that RAVE is not complete at $I_{\mathrm{2MASS}} = 13$. Quantitatively, in order to 
characterize the skewness of each distribution we compute quartile values ($Q_1, Q_2, Q_3$), which represent 
the 25\textsuperscript{th}, 50\textsuperscript{th}, and 75\textsuperscript{th} percentiles, respectively.

We investigate these potential biases in three subsamples: giants (log$g < 3.5$), the main sequence region (log$g > 4.0, T_{\mathrm{eff}} < 5500 $K), and the
turnoff region (log$g >3.5, 5500 \mathrm{K} < T_{\mathrm{eff}} < 7000$K). The boundaries of these subsamples have been determined 
from the $T_{\mathrm{eff}}$- log$ g$ plane of our parent \textsc{Galaxia} sample (see top row of Figure~\ref{fig:2dhist_compare}). For these comparisons we also consider the distance $|z|$ 
from the Galactic plane by dividing our subsamples into 3 bins of height above the plane. The size of these bins 
varies between our subsamples, as these populations probe different distance distributions. 

\begin{figure}
 \centering
 \includegraphics[width=0.49\textwidth]{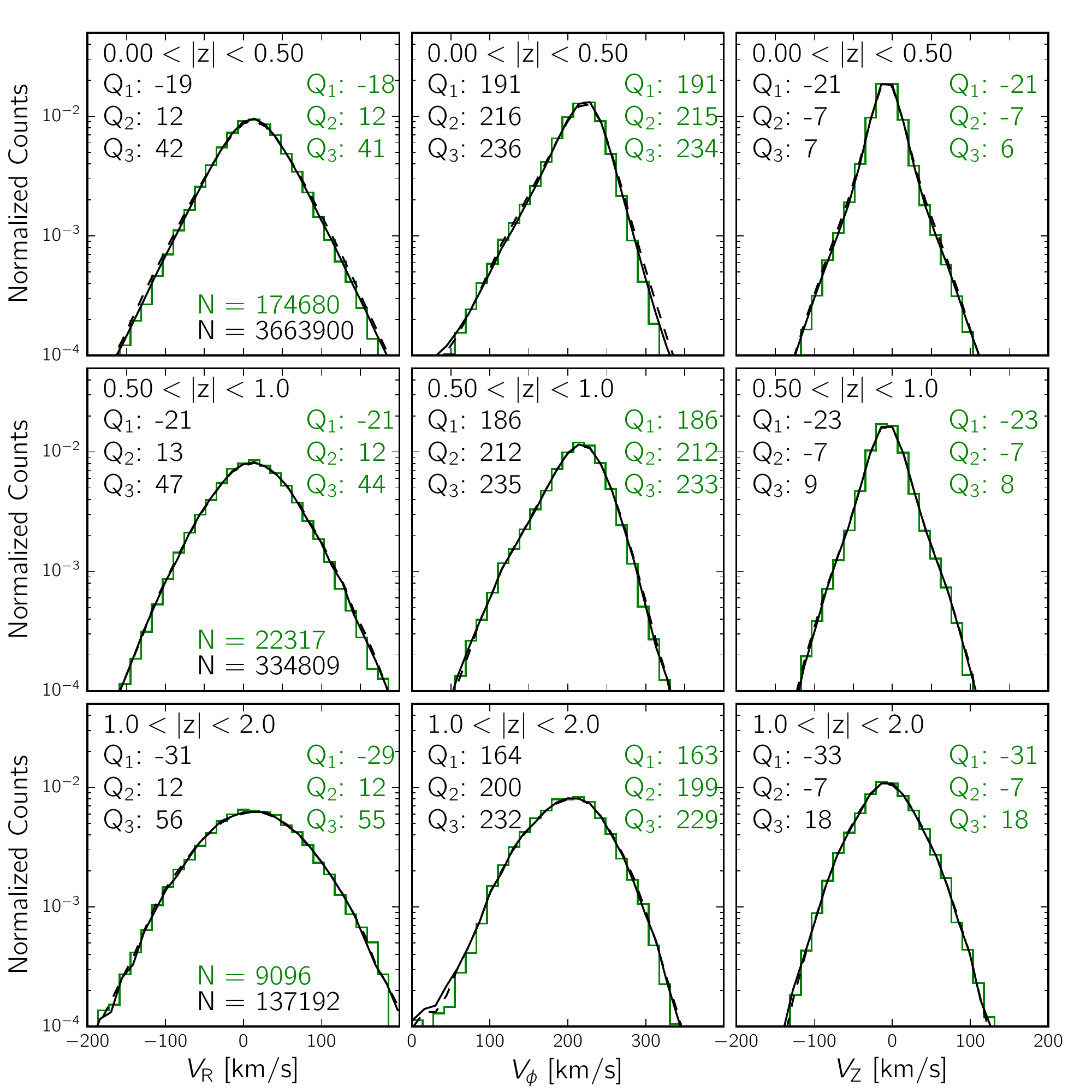}
 \caption{Distributions of Galactocentric cylindrical velocity components for samples of giant stars (log$g < 3.5$)
 at different distances from the Galactic plane as 
 indicated in the panels. The green histograms show the velocity distributions in the mock-RAVE 
 catalogue, while the black-dashed curves show the distributions for our parent \textsc{Galaxia}  
 subsample of giants. Solid black curves show the distribution for a parent \textsc{Galaxia} sample limited to $I < 12$. Quantile values ($Q_1, Q_2, Q_3$) for both distributions are given in each panel, which represent the 
 25\textsuperscript{th}, 50\textsuperscript{th}, and 75\textsuperscript{th} percentiles, respectively. {The sample 
size (N) for the distributions are shown in green and black, representing the mock-RAVE sample and the parent 
Galaxia sample limited to $I < 12$, respectively.} The y-axis is plotted on a logarithmic scale.}
 \label{fig:Impact_rv}
\end{figure}

\begin{figure}
 \centering
 \includegraphics[width=0.49\textwidth]{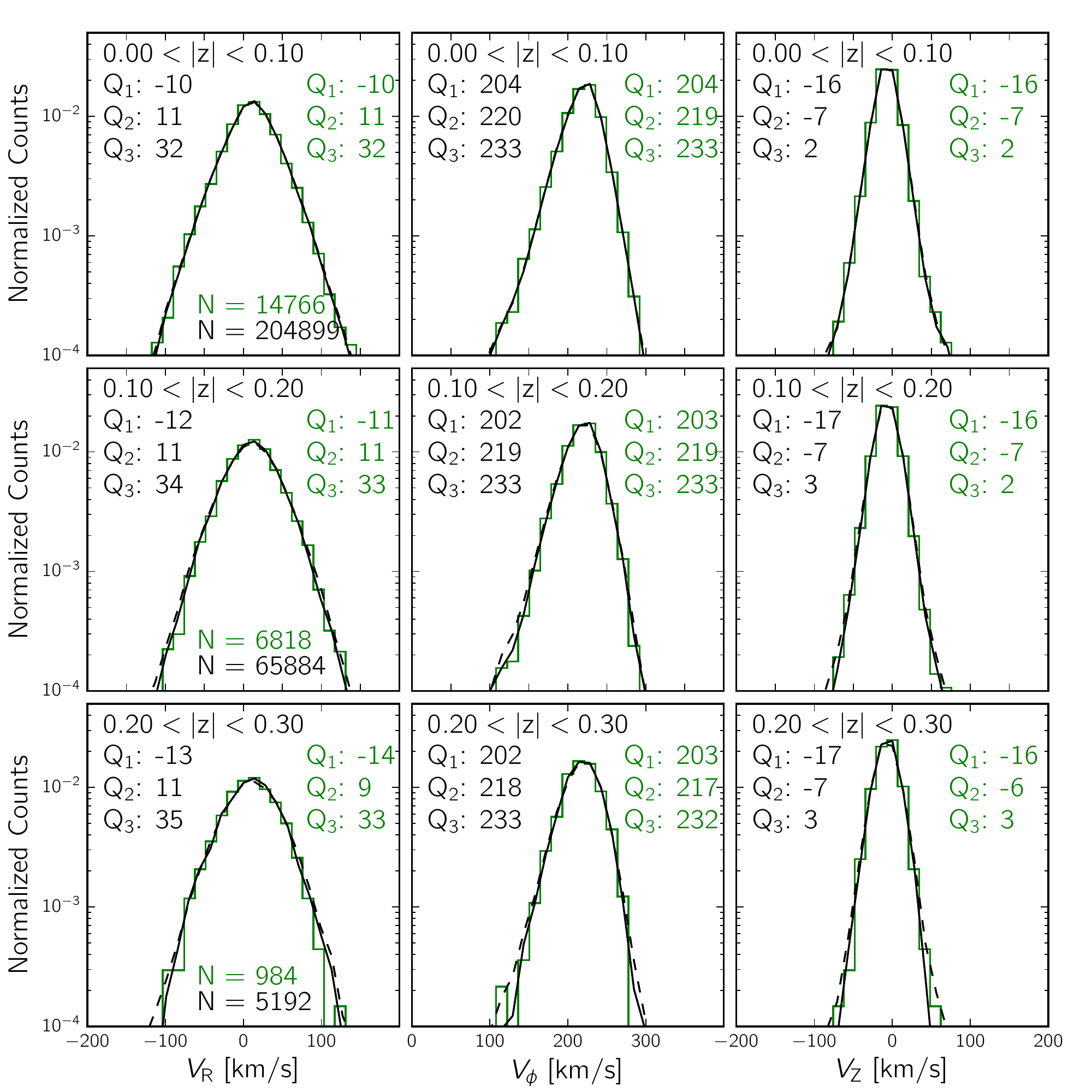}
 \caption{Same as Figure~\ref{fig:Impact_rv} but for the main sequence region (log$g~>~4.0, T_{\mathrm{eff}}~<~5500 $K) 
 sample.}
 \label{fig:Impact_rv_cdwarf}
\end{figure}

\begin{figure}
 \centering
 \includegraphics[width=0.49\textwidth]{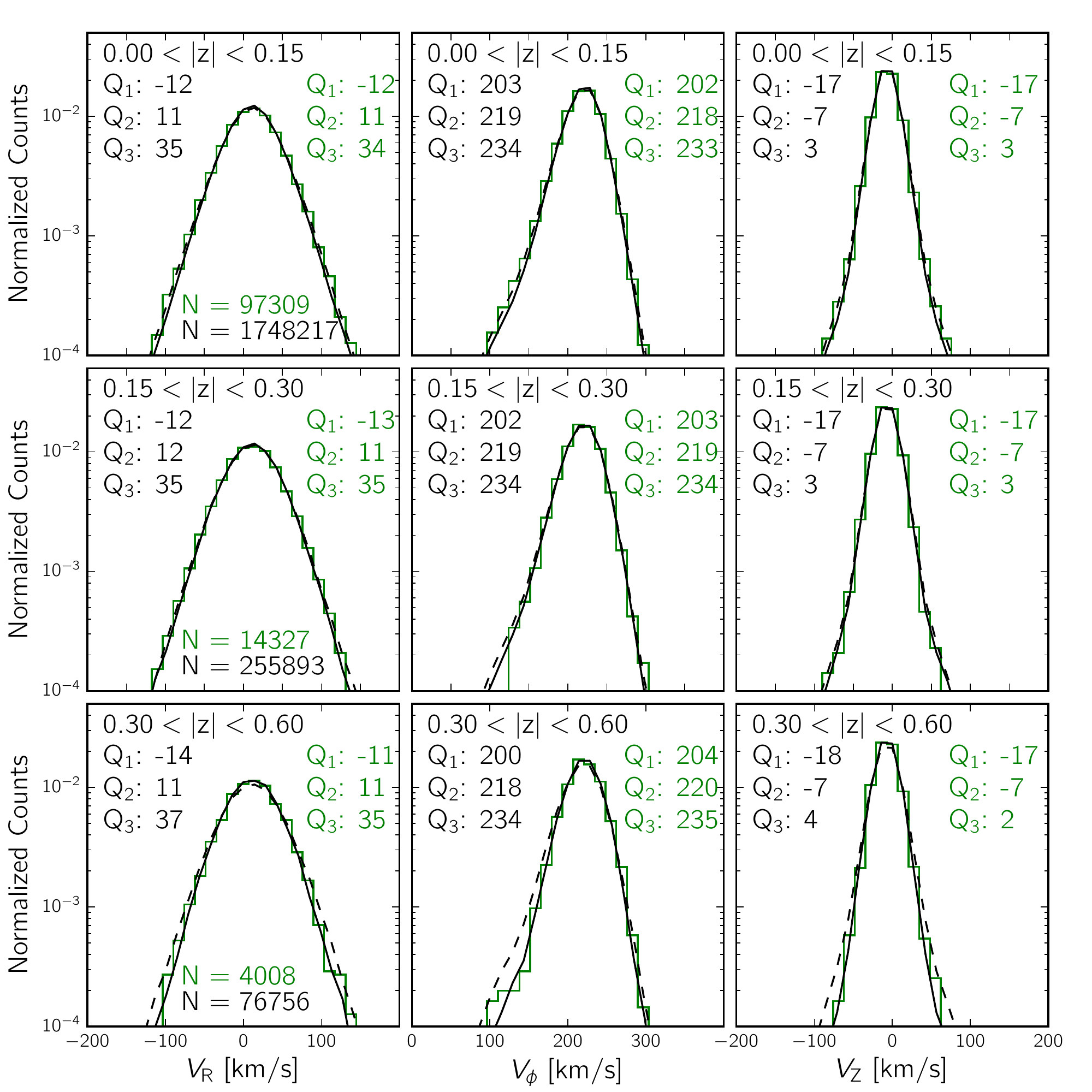}
 \caption{Same as Figure~\ref{fig:Impact_rv} but for the turnoff region (log$g~>~3.5, 5500$\,K~$< T_{\mathrm{eff}} < 7000$\,K)
  sample.}
 \label{fig:Impact_rv_hdwarf}
\end{figure}

\subsubsection{Velocity distribution comparison}
\label{sec:velocity}

We first examine the effect of our selection function on distributions of the cylindrical Galactocentric velocity 
components in our mock-RAVE catalogue. Our results are shown in 
Figures~\ref{fig:Impact_rv}, \ref{fig:Impact_rv_cdwarf}, and \ref{fig:Impact_rv_hdwarf}, with the \textsc{Galaxia} 
distribution shown as dashed black curves, and the mock-RAVE catalogue shown in green. A \textsc{Galaxia} distribution limited to $I < 12$ is shown as solid black curves. Quartile values are given in each 
panel.

For our giant and main sequence region samples (Figures~\ref{fig:Impact_rv} and \ref{fig:Impact_rv_cdwarf}), we find nearly identical distributions for all distance bins when 
comparing our mock-RAVE catalogue with the respective parent \textsc{Galaxia} distributions. We 
consider the distributions to agree if we find all three quartiles to agree within $5\kms$. Using this criterion, 
we confirm that the selection function does not impose kinematic biases for these populations as a 
whole. {We note that when we consider only low-latitude fields ($5^\circ < |b| < 25^\circ$), the colour criterion that was imposed to select preferentially for giants (see Section~\ref{sec:quality_rave}) reflects to a small bias in age. Further comparisons with the model have shown that this age bias does not introduce a significant kinematic bias, however, we urge some caution when considering the velocity distributions for these low-latitude fields.} 

We also find good agreement in most height bins for each velocity component of our turnoff region sample (Figure~\ref{fig:Impact_rv_hdwarf}). However, for the most distant bin ($0.30 < |\mathrm{z}| < 0.60$~kpc), there is a slight 
difference between the distributions in the low-$V_{\phi}$ tail. Specifically, the application of the selection 
function leads to an underrepresentation of stars with $V_{\phi} \lesssim 150\kms$ in our mock-RAVE 
catalogue. Bias is present in all components of the velocity, but we find it most clearly in $V_{\phi}$, as the 
velocity distribution functions for the thin disc, thick disc, and halo do not have the same mean for this 
component. 

The difference that we find can be explained by the difference in magnitude 
distributions between our two samples: our parent \textsc{Galaxia} sample 
extends to $I_{\mathrm{2MASS}} \sim 13$ (see Section~\ref{sec:Galaxia}), whereas our mock-RAVE sample follows the $I$-magnitude 
distribution of RAVE (see Figure~\ref{fig:RAVE_Imag}), by the definition of the selection function. As a 
consequence, there are relatively few stars observed in RAVE with $12 < I < 13$ compared to those present in 
our parent \textsc{Galaxia} sample. By having a larger fraction of stars at fainter magnitudes, the parent \textsc{Galaxia} sample 
probes more of the thick disc and halo compared to our mock-RAVE sample. This effect also is reflected in 
differences that we see between the metallicity distributions (see Section~\ref{sec:metallicity_comparison} and Figure~\ref{fig:Impact_metallicity}). This 
discrepancy is small (and indeed disappears if we limit our parent \textsc{Galaxia} sample to $I_{\mathrm{2MASS}} < 12$), and overall the distributions meet our criterion (all three quartiles agree within $5\kms$), so we 
consider the turnoff region stars to also be kinematically unbiased. 

Similar tests were done for a sample of hot dwarf stars (log$g > 3.5, T_{\mathrm{eff}} > 7000 $K), but are not shown here. As with the our turnoff region sample, we find our sample of  hot dwarfs to also be unbiased for $I < 12$.

\subsubsection{Metallicity distribution comparison}
\label{sec:metallicity_comparison}

\begin{figure}
 \centering
 \includegraphics[width=0.49\textwidth]{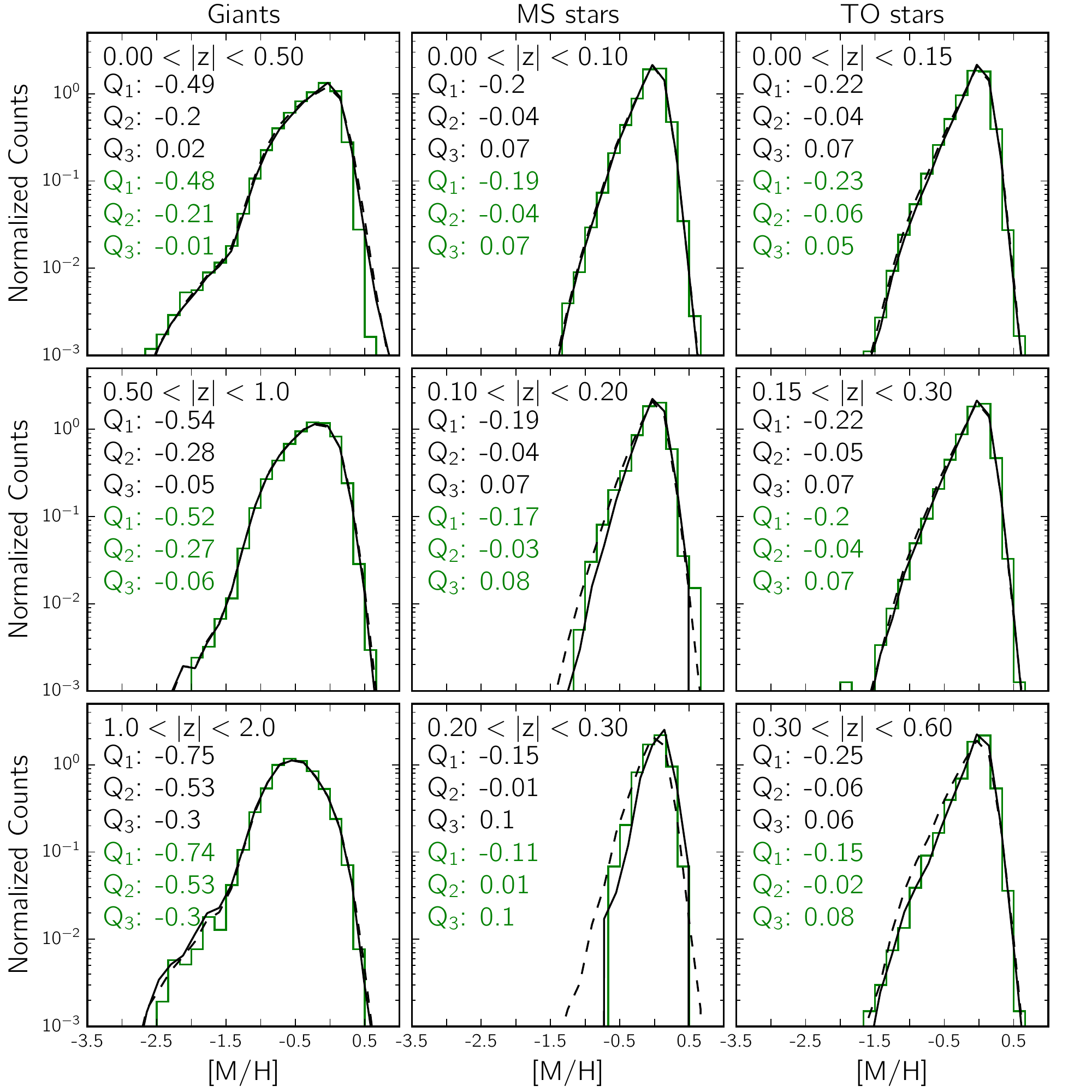}
 \caption{Metallicity distributions for each sample in different distances from the Galactic plane as indicated in the panels. The left column shows the distributions for giants, the middle shows main sequence region stars, and the right shows turnoff region stars. The black-dashed curves indicate the underlying distributions for our \textsc{Galaxia} parent sample, while the green histograms show the metallicity distributions in the mock-RAVE catalogue. Solid black curves show the distribution for a parent \textsc{Galaxia} sample limited to $I < 12$. Quartile values ($Q_1, Q_2, Q_3$) for both distributions are given in each panel. The y-axis is plotted on a logarithmic scale.}
 \label{fig:Impact_metallicity}
\end{figure}

Next, we examine the metallicity distributions of the \textsc{Galaxia} samples and our mock-RAVE 
catalogue. The metallicity distributions for each subsample in different slices in distance $|z|$ from the Galactic 
plane are shown in Figure~\ref{fig:Impact_metallicity}. Here, we consider the distributions to agree if all three 
quartiles agree within $0.1$ dex.

For giants (left column of Figure~\ref{fig:Impact_metallicity}) and stars in the main sequence region (middle column of Figure~\ref{fig:Impact_metallicity}), we find very good agreement between the \textsc{Galaxia} and mock-RAVE metallicity 
distributions for all distance bins. For stars in the main sequence region and the most distant bin ($0.20 < |\mathrm{z}| < 0.30$ kpc), we find that in our mock-RAVE sample the metal-poor tail of the 
metallicity distribution is slightly underrepresented, compared to the \textsc{Galaxia} sample. However, this difference can be explained by small 
number statistics, as our mock-RAVE sample would need only one star below [M/H]~$\sim-0.6$ to reconcile the 
difference between the two distributions. Again, despite this small discrepancy, the quartile values satisfy our 
criterion, and therefore we consider our main sequence region sample to be chemically unbiased. We  
conclude that for giants and stars in the main sequence region, our metallicity distribution is minimally 
affected by our selection function.

Similarly, for the turnoff region sample (right column of Figure~\ref{fig:Impact_metallicity}), we find good 
agreement for the two closest distance bins, with differences between the two distributions found only in the 
furthest distance bin ($0.30 < |\mathrm{z}| < 0.60$ kpc). For this bin, we find that our criterion is barely met, 
with $Q_1$ differing by $\sim0.1$ dex. This discrepancy between the two distributions is explained by the 
difference in magnitude limits as described in Section~\ref{sec:velocity}. That is, as our parent \textsc{Galaxia} sample 
includes a larger fraction of faint ($12 < I < 13$) stars compared to our mock-RAVE sample, it probes a larger 
volume, and therefore more of the thick disc and halo. This effect is less prominent for our giant sample, as the 
relative fractional increase of thick disc and halo stars is much less for giants, compared to our dwarf sample. 
We conclude that our turnoff region sample is unbiased for $I_{\mathrm{2MASS}} < 12$. 
As with the velocity comparisons, we also test the [M/H] distributions for a sample of hot dwarf stars (log$g > 3.5, T_{\mathrm{eff}} > 7000$K), and find them to also be chemically unbiased.

\section{Discussion and conclusions}
\label{sec:conclusion}
We have described, in detail, how to evaluate the selection function $S$ of the RAVE survey in two different 
ways: field-by-field, and \textsc{healpix} pixels. In addition, we discussed the uncertainty distributions of  
RAVE DR5 and illustrated that these uncertainties depend heavily on the position in stellar parameter space. 
We then generated a mock-RAVE catalogue by applying the detailed selection function to the model output, 
and modified the raw \textsc{Galaxia} output by RAVE-like {uncertainties}.

To investigate that RAVE is a kinematically and chemically unbiased survey, we tested the impact of $S$ on the 
resulting velocity and metallicity distributions using a modified version of the Besan\c{c}on model available in 
the \textsc{Galaxia} framework. The velocity and metallicity distributions of our mock-RAVE catalogue were compared  
with the distributions of the underlying \textsc{Galaxia} populations. {We find that, for $I < 12$, our 
selection function does not intrinsically induce biases in the kinematics or chemistry of stars within the stellar 
parameter space covered in RAVE ($4000 \rm K < T_{\rm eff} < 8000 \rm K$ and $0.5 < \mathrm{log}\,g < 5.0$), 
with respect to expectations from the Besan\c{c}on model available in Galaxia. We do find some small biases 
when we consider a parent sample extending to $I =13$, however, it has been shown that the completeness of 
RAVE falls off for fainter magnitudes (due to the magnitude limit imposed from the input catalogues), and 
therefore our conclusion stands for the magnitude range where we consider RAVE to provide a representative 
sample of stars ($ 9 < I < 12$). Under these criteria, and within this parameter space, RAVE stars provide unbiased samples in terms of 
kinematics and metallicities that are well suited for kinematic modeling 
without taking into account the detailed selection function via volume corrections.}

For our giant and main 
sequence region samples, we find good agreement between the parent \textsc{Galaxia} sample and our mock-RAVE 
catalogue. We find similar trends for our sample of turnoff region stars, with small differences in the velocity  
distributions for the most distant stars, and the metal-poor tail of the [M/H] distribution. However, we 
explain this bias due to the fact that our \textsc{Galaxia} sample 
includes a larger number of stars at fainter magnitudes compared to our mock-RAVE catalogue. The parent 
\textsc{Galaxia} sample therefore probes a larger volume than our mock-RAVE catalogue, and consequently more of 
the thick disc and halo populations. As we are able to account for the source of these differences, we consider 
our turnoff region sample to also be kinematically and chemically unbiased for $I_{\mathrm{2MASS}} < 12$. 

Recently, a number of studies used RAVE data, and in particular subsamples of giant stars, for kinematic 
modeling \citep[e.g.][]{Binney14, Piffl14, Minchev14, Williams13, Kordopatis13_b, Bienayme14}. Here we confirm that the giant stars in RAVE can indeed be used as an unbiased sample.
\citet{Piffl14} fitted a full dynamical model of the Milky Way to the kinematics of the RAVE giants. They then 
tested if the resulting model would also correctly predict the kinematics of a sample of hot dwarf stars from 
RAVE and found a number of discrepancies. Their conclusion was that the thick disc distribution function in 
their model was too simplistic. However, \citet{Binney14} also 
found that a similar dynamical model fitted to data from the GCS \citep{Nordstroem04} could reproduce the 
RAVE hot dwarf kinematics, but did not fit the RAVE giants. Since the GCS has a selection function that is 
different from that of the RAVE dwarfs, this implies that taking into account a more complicated volume 
correction for the hot dwarfs will not be enough to completely reconcile them with the model of \citet{Piffl14}. 
Hence a more complex distribution function for the thick disc, as argued for by the authors, seems still 
necessary.

We also illustrate that the quantified RAVE selection function can be used to generate mock-RAVE surveys 
from stellar population synthesis models, and in combination with code frameworks like \textsc{Galaxia}, it can 
serve as a powerful tool to test Galaxy models against the RAVE data. {The two versions of the RAVE selection 
function produced by this study (field-by-field and by \textsc{healpix} pixel) will be made publicly available on the RAVE web site (\href{https://www.rave-survey.org}{https://www.rave-survey.org}).}

\section*{Acknowledgements}
JW thanks Ivan Minchev, Friedrich Anders, Else Starkenburg, Kris Youakim, and Alexey Mints for their comments and 
helpful discussions, which have improved the quality and clarity of the text.
Funding for this work and for RAVE has been provided by: the
Australian Astronomical Observatory; the Leibniz-Institut
fuer Astrophysik Potsdam (AIP); the Australian National
University; the Australian Research Council; the European Research Council
under the European Union's Seventh Framework
Programme (Grant Agreement 240271 and 321067); the French National
Research Agency; the German Research Foundation
(SPP 1177 and SFB 881); the Istituto Nazionale di
Astrofisica at Padova; The Johns Hopkins University; the
National Science Foundation of the USA (AST-0908326);
the W. M. Keck foundation; the Macquarie University;
the Netherlands Research School for Astronomy; the Natural
Sciences and Engineering Research Council of Canada;
the Slovenian Research Agency (research core funding No. P1-0188); the Swiss National Science
Foundation; the Science \& Technology Facilities Council of
the UK; Opticon; Strasbourg Observatory; and the Universities
of Groningen, Heidelberg and Sydney. The RAVE web
site is at https://www.rave-survey.org.




\bibliographystyle{mnras}
\bibliography{mybib} 







\bsp	
\label{lastpage}
\end{document}